\documentclass[aps,eqsecnum,preprint,floats,epsf,epsfig,nofootinbib,showpacs]{revtex4}
\usepackage{graphicx}
\usepackage{amsmath}
\def\be{\begin{eqnarray}}
\def\en{\end{eqnarray}}
\def\non{\nonumber}
\def\la{\langle}
\def\ra{\rangle}
\def\pl{{ Phys. Lett.}~}
\def\pr{{ Phys. Rev.}~}

\def\bi{\bibitem}
\begin{document}

\title{\Large \bf Strong and radiative decays of heavy mesons in a covariant model}

\author{ \bf Chi-Yee Cheung$^a$ and Chien-Wen Hwang$^b$\footnote{
t2732@nknucc.nknu.edu.tw}}
\vskip 1.4 cm
\affiliation{\centerline{$^a$ Institute of Physics, Academia Sinica, Taipei, Taiwan 115, Republic of China}\\
\centerline{$^b$ Department of Physics, National Kaohsiung Normal University, Kaohsiung,} \\
\centerline{Taiwan 824, Republic of China}}


\begin{abstract}
In this paper, we investigate symmetry breaking effects in strong and radiative decays of heavy mesons.
We study $1/m_Q$ corrections within the heavy quark effective theory. These effects are studied in a covariant model for heavy mesons. The numerical results are consistent with the experimental data and some other theoretical calculations.
These provide a vote of confidence for the validity of this covariant model.

\end{abstract}
\pacs{13.20.Fc, 12.39.Ki, 13.20.Fc, 13.20.He}
\maketitle %
\section{Introduction}
Recently, the BaBar Collaboration \cite{data1,data2} measured the mass difference between the $D^*(2010)^+$ and the $D^0$ and the natural line width of the $D^*(2010)^+$ vector meson; they obtained very precise values: $\Delta m_{D^{*+}D^0}=(145425.9 \pm 0.4 \pm 1.7)$ keV and $\Gamma_{\textrm{tot}}(D^{*+}) = (83.3\pm1.2\pm1.4)$ keV. For $D^*$ mesons, the strong pion emission is one of the dominant decay modes which determine their life times. Thus, these precise values provide an ideal occasion to test different theoretical estimations for the strong interactions of the heavy mesons.

It is widely accepted that quantum chromodynamics (QCD) is the correct theory for strong interactions. QCD is a renormalizable quantum field theory which is closely modeled after quantum electrodynamics (QED), the most accurate physical theory we have to date. In QCD, the colored quarks interact by exchanging $SU(3)$ Yang-Mill gauge fields (gluons) which are also colored, so that gluons interact directly among themselves. This feature leads to the strength of the quark-gluon interaction decreases with increasing momentum transfers (asymptotic freedom) \cite{GW,P}. Thus, in high-energy processes, quarks and gluons only weakly interact, and physical processes become describable by the perturbation theory. Numerous experiments performed on powerful particle accelerators around the world have attested to the extraordinary success of this theory at high energy \cite{PDG12}. On the other hand, in the low-energy (or long-distance) regime, QCD tells us that the interactions between quarks and gluons are strong, so that quark-gluon dynamics becomes nonperturbative in nature. Consequently, for low energy phenomena, it is hard to perform any reliable calculations from first principles. Thus, understanding the structures of hadrons directly from QCD remains an outstanding problem. In 1989, it was realized that, in low energy situations where the typical gluon momenta are small compared with the heavy quark mass $(m_Q)$, QCD dynamics becomes independent of the flavor (mass) and spin of the heavy quark \cite{iw,Geo1}. These new spin and flavor symmetries combine to form a $SU(2N^Q_f)$ symmetry, called heavy quark symmetry (HQS), which is not manifest in the original QCD Lagrangian. Of course, even in this infinite heavy quark mass limit, low energy QCD dynamics remains nonperturbative, and there are still no solutions to old problems like quark confinement, etc. HQS allows us to factorize the complicated light quark and gluon dynamics from that of the heavy one, and thus provides a clearer physical picture in the study of heavy quark physics. Beyond the symmetry limit, a heavy quark effective theory (HQET) can be developed by systematically expanding the QCD Lagrangian in powers of $1/m_Q$, with which HQS breaking effects can be studied order by order \cite{Geo1,NN,MRR}.

In the other extreme, due to the relatively small light quark masses $(m_u, m_d, m_s)$, the light quark sector of the QCD Lagrangian obeys an approximate $SU(3)_L \times SU(3)_R$ chiral symmetry \cite{DGH}. It was known early on that this symmetry must be spontaneously broken by the QCD vacuum, so that the physical spectra of hadrons made up of light quarks would have only $SU(3)_{L+R}$ symmetry. Moreover, due to the spontaneous breaking of the chiral symmetry, there exist eight pseudoscalar bosons (called Goldstone bosons, which include three $\pi$'s, four $K$'s, and one $\eta$), whose dynamics obeys the $SU(3)_L \times SU(3)_R$ chiral symmetry. From the above discussions, it should be clear that if we want to study the low energy interactions of heavy hadrons and Goldstone bosons, we need to build an effective theory that obeys both chiral and heavy quark symmetries. This is done in references \cite{Wise92,BD92,HYC1,HYC3,HYC4}, where chiral symmetry and HQS are synthesized in a single effective chiral Lagrangian which describes the strong interactions between heavy hadrons and Goldstone bosons. The theory has since been extended to incorporate electromagnetic interactions as well \cite{CG,Amundson,HYC2,HYC3,HYC4}.

For excited heavy mesons $(D^*,B^*)$, pion and photon emissions are the dominant decay modes which determine their life times \cite{PDG12}. In principle, the effective chiral Lagrangian provides an ideal framework in which to study the strong decay mode. However, symmetry considerations alone in general do not lead to quantitative predictions, unless further assumptions are made to extract the values of the various coupling constants appearing in the Lagrangian. Furthermore, the framework of an effective chiral Lagrangian does not allow for a systematic discussion of HQS violating $1/m_Q$ effects, which however is important for a thorough understanding of heavy quark physics. 
The purpose of this paper is to systematically study the $1/m_Q$ effects in a covariant model for strong and radiative decays of heavy meson.

The paper is organized as follows. In Sec. II, we briefly review HQET and the chiral dynamics of heavy meson. In Sec. III, we construct a covariant model which is based on HQET. Some heavy meson properties in the heavy quark limit and $1/m_Q$ corrections are considered. The numerical calculations and discussions are expressed in Sec. IV. In Sec. V, we make some concluding remarks.

\section{Brief Reviews}
\subsection{HQET}
The full QCD Lagrangian for a heavy quark ($c$, $b$, or $t$) is given by
\be
   L_Q = \bar Q~(i\gamma_\mu D^\mu - m_Q)~Q,   \label{Lag}
\en
where $D^\mu \equiv \partial ^\mu - i g_s T^a A^{a\mu}$ with $T^a = \lambda^a/2$. The heavy quark $Q$ interacts with the light degrees of freedom by exchanging gluons with momenta of order $\Lambda_{\textrm{QCD}}$, which is much smaller than its mass $m_Q$. Thus, the heavy quark's momentum $p_Q$ is close to the ``kinetic" momentum $m_Q v$ resulting from the meson's motion:
 \be
   p^\mu_Q = m_Q v^\mu + k^\mu,  \label{Pk}
 \en
where $k^\mu$ is the so-called ``residual" momentum and is of order $\Lambda_{\textrm{QCD}}$. In the limit $m_Q \to \infty$, one can define a new heavy quark field $h_v (x)$, which is related to the original field $Q(x)$ by
 \be
   {1+\not\!v \over {2}}Q(x) = e^{-im_Q v\cdot x}~h_v (x), \label{move}
 \en
where $h_v (x)$ satisfies the constraints
 \be
   {1+\not\!v \over {2}} h_v = h_v,\qquad i\partial^\mu h_v(x) = k^\mu h_v (x).
 \en
Therefore, the heavy quark QCD Lagrangian, Eq. (\ref{Lag}), is reduced to:
\be
   L_Q \to L_{Q,\textrm{eff}}= \bar h_v iv\cdot D h_v. \label{L00}
\en
From Eq. (\ref{L00}), it is evident that this effective Lagrangian possesses {\it flavor} and {\it spin} symmetries, also known as heavy quark symmetry (HQS), because it is independent of the heavy quark mass $m_Q$ and the $\vec \gamma$ matrix, respectively. Thus, HQS predicts that for the $Q\bar q$ system, the pseudoscalar $(0^-)$ and the vector $(1^-)$ states are degenerate. Experimentally, the pseudoscalar-vector (or hyperfine) mass splitting $\delta m_{_{HF}}$ is given by \cite{PDG12}:
\be
   m_{B^*} - m_B \approx 45.78~\textrm{MeV},  \non \\
   m_{D^*} - m_D \approx 142.12~\textrm{MeV},     \label{HQSmass}
\en
from which we see that the experimental mass difference gets smaller as the mass of the heavy quark gets heavier. This result inspires us with the theory that the corrections to heavy quark symmetry are of first order in ${\cal O} (1/m_Q)$.
In heavy quark effective theory \cite{Geo1,NN}, the QCD Lagrangian is expanded as:
\be
   L_{Q,\textrm{eff}} &=& \bar h_v i v\cdot D h_v + \sum_{n=1}^{\infty}\left(\frac{1}{2 m_Q}\right)^n\bar h_v i \not\!\!D_{\bot} (-i v \cdot\! D)^{n-1} i \not\!\!D_\bot h_v, \non \\
   &=& \bar h_v i v\cdot D h_v + {1\over {2 m_Q}} \bar h_v (iD_\bot)^2 h_v + {g\over {4 m_Q}} \bar h_v \sigma_{\alpha\beta} G^{\alpha\beta} h_v + {\cal O} \left({1\over {m^2_Q}}\right), \label{expand}
\en
where $D^\mu_{\bot} = D^\mu - v^\mu v \cdot D$ is orthogonal to the heavy quark velocity and $G^{\alpha\beta} = T_a G^{\alpha\beta}_a = {i\over {g_s}}[D^\alpha,D^\beta]$ is the gluon field strength tensor. This is the generalization of Eq. (\ref{L00}) to finite heavy quark mass and the new operators at order $1/m_Q$ are:
\be
   {\cal O}_1 = {1\over {2 m_Q}}~{\bar h}_v~(iD_\bot)^2~ h_v,  \label{O1} \\
  {\cal O}_2 = {g\over {4 m_Q}}~{\bar h}_v~\sigma^{\mu\nu}~G_{\mu\nu}~h_v, \label{O2}
\en
where ${\cal O}_1$ is the gauge invariant extension of the kinetic energy arising from the off-shell residual motion of the heavy quark, and ${\cal O}_2$ describes the color magnetic interaction of the heavy quark spin with the gluon field. It is clear that both ${\cal O}_1$ and ${\cal O}_2$ break the flavor symmetry, while ${\cal O}_2$ breaks the spin symmetry.
We note that, HQS is a symmetry of the lowest order of $L_{Q,\textrm{eff}}$, therefore the predictions from HQS are model independent. However, until the QCD bound state problem is solved, quantitative effects of the higher order of $L_{Q,\textrm{eff}}$ would have to be evaluated in specific models for hadrons.
\subsection{Chiral Lagrangians for strong decays}
The QCD Lagrangian for light quarks is:
\be
   {\cal L}_q = \bar q~(i\gamma_\mu D^\mu - m_q)~q,   \label{Lagq}
\en
where $q=(u,d,s)$. In the limit $m_q \to 0$, ${\cal L}_q$ possesses an $SU(3)_L \times SU(3)_R$ flavor chiral symmetry. That means ${\cal L}_q$ is invariant under the independent transformations
$q_L \to L q_L$ and $q_R \to R q_R$ 
are the left-handed and right-handed quark fields, respectively, and $L(R)=e^{i\theta^a_{L(R)} T^a}$ is the global transformations in $SU(3)_{L(R)}$. 
By Goldstone theorem, the spontaneous symmetry breaking of the QCD vacuum generates eight Goldstone bosons: $(\pi^{+,-,0}, K^{+,-,0}, \bar K^0$ and $\eta)$, whose dynamics is still governed by the full $SU(3)_L \times SU(3)_R$ symmetry. Since the heavy mesons contain both heavy and light quarks, one expects that when we study the low-energy interactions of heavy mesons with the Goldstone bosons, both the chiral symmetry and HQS will play important roles.
\def\pke%
{\begin{array}{ccc}
{\pi^0 \over {\sqrt{2}}} + {\eta \over {\sqrt{6}}} & \pi^+ & K^+ \\
\pi^- & -{\pi^0 \over {\sqrt{2}}} + {\eta \over {\sqrt{6}}} & K^0 \\
K^- & \bar K^0 & -\sqrt {2 \over {3}} \eta
\end{array}}

The low lying states of a $(Q \bar q)$ system consist of pseudoscalar $(0^-)$ and vector $(1^-)$ mesons, which we denote by $P$ and $V$, respectively. In the HQS limit, their quantum numbers can be conveniently incorporated in the interpolating fields \cite{Geo}:
\be
   P(v) &=& i\bar q \gamma_5 h_v \sqrt {M_H}, \label{Pin}\\
   V(v,\epsilon) &=& \bar q \not\! \epsilon h_v \sqrt {M_H}, \label{Vin}
\en
where $M_H$ is the heavy meson mass and $h_v$ is defined in Eq. (\ref{move}). Thus, both $P(v)$ and $V(v,\epsilon)$ are flavor $SU(3)$ antitriplets. Before discussing the interaction between heavy mesons and Goldstone bosons, we will first summarize the case of Goldstone bosons interacting among themselves \cite{HYC1,Geob}. The chiral symmetry is nonlinearly realized by using the unitary matrix $\Sigma = e^{2iM/\sqrt{2}f_\pi}$, $M$ is a $3 \times 3$ matrix for the octet of Goldstone bosons
\be
  M= \left[
     \pke
     \right]
\en
and $f_\pi = 93$ MeV is the pion decay constant. Under $SU(3)_L \times SU(3)_R$, $\Sigma$ transforms as
$\Sigma \to \Sigma' = L \Sigma R^\dagger$.
In order to facilitate the discussions of the Goldstone bosons interacting with heavy mesons, we introduce a new Goldstone-boson matrix $\xi \equiv \Sigma^{1/2}$, which transforms under an $SU(3)_L \times SU(3)_R$ as:
$\xi \to \xi' = L~\xi~U^\dagger = U~\xi~R^\dagger$, where $U$ is a unitary matrix depending on $L$, $R$, and $M$, so that it is no longer global. Now with the aid of $\xi$, we construct a vector field ${\cal V}_\mu$ and an axial vector field ${\cal A}_\mu$:
\be
   {\cal V}_\mu &=& {1\over {2}}(\xi^\dagger \partial_\mu \xi + \xi \partial_\mu \xi^\dagger), \label{vector}\\
   {\cal A}_\mu &=& {i\over {2}}(\xi^\dagger \partial_\mu \xi - \xi \partial_\mu \xi^\dagger), \label{axial}
\en
with the simple transformation properties:
\be
   {\cal V}_\mu \to {\cal V}'_\mu &=& U~{\cal V}_\mu~U^\dagger + U~\partial_\mu~U^\dagger, \label{Vec}\\
   {\cal A}_\mu \to {\cal A}'_\mu &=& U~{\cal A}_\mu~U^\dagger.
\en
The light quarks can be made to transform simply as:
\be
   q \to q' = U~q.  \label{qU}
\en
Now we can study the chiral properties of the heavy mesons. According to Eqs. (\ref{Pin}) and (\ref{qU}), the flavor antitriplet $P$ transforms as:
\be
   P \to P' = P~U^\dagger, \label{Pt}
\en
Using the vector field ${\cal V}_\mu$ in Eq. (\ref{Vec}), we can construct a covariant derivative $D'_\mu$:
\be
   {\cal D}_\mu P^\dagger \equiv (\partial_\mu + {\cal V}_\mu)P^\dagger  \label{Dpd}
\en
which transforms simply as:
\be
   {\cal D}_\mu P^\dagger \to ({\cal D}_\mu P^\dagger)' = U({\cal D}_\mu P^\dagger),\label{Dpt}
\en
and similarly for vector mesons $V$.

We can now use Eqs. (\ref{Pt} $-$ \ref{Dpt}) to construct an effective Lagrangian of $P$ and $V$ and their couplings to the Goldstone bosons \cite{HYC1}:
\be
   L_{VP} &=& {\cal D}_\mu P~{\cal D}^\mu P^\dagger - M^2_H PP^\dagger + if M_H (P~{\cal A}^\mu V_\mu^\dagger - V_\mu {\cal A}^\mu P^\dagger)-{1\over {2}} V^{\mu\nu}V^\dagger_{\mu\nu}  \non \\
&+& M^2_H V^\mu V_\mu^\dagger+{1\over {2}}g \epsilon _{\mu\nu\alpha\beta}(V^{\mu\nu}{\cal A}^\alpha V^{\beta\dagger} + V^\beta {\cal A}^\alpha V^{\mu\nu\dagger}), \label{LVP}
\en
where $V^\dagger_{\mu\nu} = {\cal D}_\mu V_\nu^\dagger - {\cal D}_\nu V_\mu^\dagger$.
This is the most general Lagrangian consistent with the chiral invariance that has only one single
derivative on the Goldstone boson fields. The coupling constants $f$ and $g$ are illustrated in
Figure $1$. In Eq. (\ref{LVP}), heavy quark flavor symmetry dictates that $f$ and $g$ are universal,
i.e., they are independent of the heavy quark species involved. Furthermore, using the interpolating
field given in Eqs. (\ref{Pin} $-$ \ref{Vin}), one can easily prove that (see \cite{HYC1} for details) $f=2g$,
which is a consequence of heavy quark spin symmetry.
\begin{figure}
 \includegraphics*[width=5in]{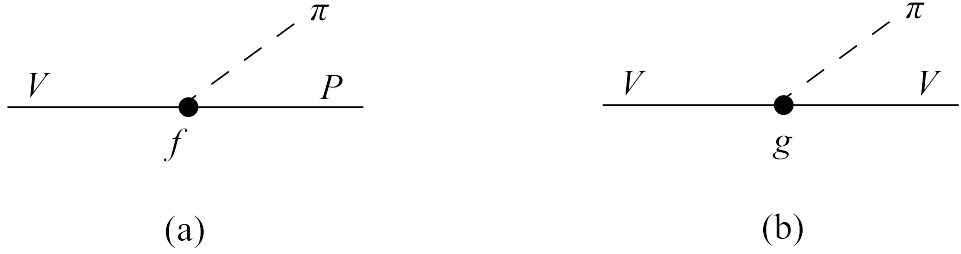}
 \caption{Strong coupling constants $f$ and $g$.}
  \label{fig:1}
 \end{figure}
\subsection{Chiral Lagrangians for radiative decays}
\def\udsq%
{\begin{array}{ccc}
{2\over {3}} & 0 & 0 \\
0 & {-1\over {3}} & 0 \\
0 & 0 & {-1\over {3}}
\end{array}}
In the $(b \bar q)$ system, due to the small mass difference $m_{B*} - m_B$, $B^* \to B \gamma$ is the dominant decay channel. For $(c \bar q)$, $D^* \to D \gamma$ can be as important as $D^* \to D \pi$, because the available phase space is much larger for the former decay mode. The light quark charge is given by
 \be
  {\cal Q}= \left[
     \udsq
     \right],
 \en
while that of the heavy quark $Q$ is denoted by ${\cal Q}'$. The electromagnetic or $U(1)$ gauge transformation of the vector potential (or photon field) $A_\mu$ is:
\be
   A_\mu \to A'_\mu = A_\mu - {1\over {e}}\partial_\mu \lambda
\en
where $\lambda$ is a $U(1)$ gauge parameter. The quark fields transform as $q \to q' = e^{i{\cal Q}\lambda} q,~~~Q \to Q' = e^{i{\cal Q}'\lambda} Q$. Since the Goldstone-boson matrix $M$ is constructed from a $q$ and a $\bar q$, it follows as $M \to M' = e^{i{\cal Q}\lambda} M e^{-i{\cal Q}\lambda}$. Thus the meson field $\xi$ also transforms simply as $\xi \to \xi' = e^{i{\cal Q}\lambda} \xi e^{-i{\cal Q}\lambda}$. A covariant derivative on the field $\xi$ has the form:
\be
   D_\mu \xi = \partial_\mu \xi + ie A_\mu [{\cal Q},\xi],
\en
with the gauge transformation $ D_\mu \xi \to (D_\mu \xi)' = e^{i{\cal Q}\lambda}~D_\mu \xi~e^{-i{\cal Q}\lambda}$. In the presence of electromagnetic interactions, the vector and axial-vector currents of Eqs. (\ref{vector}) and (\ref{axial}) must be generalized to:
\be
   {\cal V}_\mu &=& {1\over {2}}(\xi^\dagger D_\mu \xi + \xi D_\mu \xi^\dagger), \label{vectorem}\\
   {\cal A}_\mu &=& {i\over {2}}(\xi^\dagger D_\mu \xi - \xi D_\mu \xi^\dagger). \label{axialem}
\en

The heavy mesons $P$ and $V$ are composed of a heavy quark $Q$ and a light antiquark $\bar q$. Therefore, they transform as:
\be
   P &\to& P' = e^{i{\cal Q}'\lambda}~P~e^{-i{\cal Q}\lambda},\\
   V &\to& V' = e^{i{\cal Q}'\lambda}~V~e^{-i{\cal Q}\lambda}.
\en
The covariant derivative on $P$ is:
\be
   D_\mu P= \partial_\mu P +ie A_\mu ({\cal Q}'P-P{\cal Q}),
\en
which transforms as:
\be
   D_\mu P &\to& (D_\mu P)' = e^{i{\cal Q}'\lambda}~D_\mu P~e^{-i{\cal Q}\lambda}.
\en
When the chiral field is included, the covariant derivative finally reads as:
\be
   D_\mu P = \partial_\mu P + {\cal V}^\dagger_\mu P + ie A_\mu ({\cal Q}'P-P{\cal Q}).
\en
Again, the corresponding equations for the vector meson $V$ are similar.

If we are interested only in the $M1$ radiative transitions $V \to P \gamma$ and $V \to V \gamma$, the relevant lowest-order chiral and gauge invariant Lagrangian is given by \cite{HYC2}:
\be
   {\cal L}^{(2)}_{VP} &=& M_H~\epsilon_{\mu\nu\alpha\beta}v^\alpha V^\beta \times \left[{1\over {2}}d(\xi^\dagger {\cal Q}\xi+\xi {\cal Q}\xi^\dagger) \right]F^{\mu\nu}P^\dagger + {\rm h.c.} \non \\
&&+iM_H F_{\mu\nu} V^{\nu}\left[-d''{1\over {2}}(\xi^\dagger {\cal Q}\xi+\xi {\cal Q}\xi^\dagger)\right]V^{\mu\dagger}.\label{4d}
\en
where $F_{\mu\nu} = \partial_{\mu} A_\nu - \partial_{\nu} A_\mu$. The coupling constant $d$ is the transition magnetic moment for the $V \to P \gamma$ process, while $d''$ corresponds to the magnetic moment of the heavy vector meson $V$. These magnetic coupling constants are illustrated in Figure $1.3$. HQS implies that they are universal (i.e., independent of heavy quark flavors). Again, using the interpolating fields of heavy mesons, we can easily show that \cite{HYC2}:
 \be
   d'' = 2d. \label{d2d}
 \en
\begin{figure}
 \includegraphics*[width=5in]{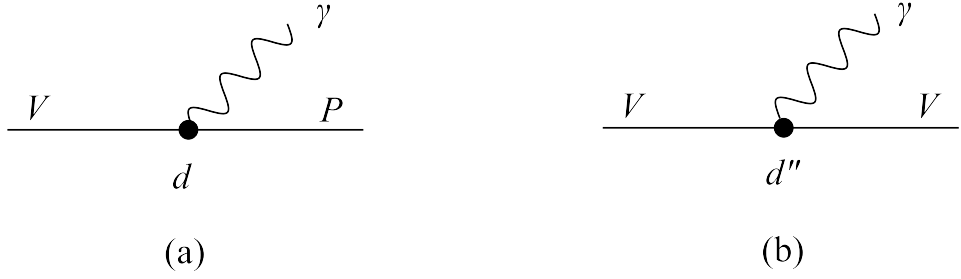}
 \caption{Magnetic coupling constants $d$ and $d"$.}
  \label{fig:d}
 \end{figure}
Note that in \cite{HYC2}, the corresponding relation is $d''=-2d$; this distinction comes from the different phase convention used in this paper. Furthermore, ${\cal L}^{(2)}_{VP}$ of Eq. (\ref{4d}) is compatible with HQS, whereas in \cite{HYC2}, some $1/m_Q$ effects are also included. For the case of $D^* \to D \gamma$, we would expect that a contribution from the charm quark magnetic moment is important since it carries a charge of ${2\over {3}}$ and its mass is not too large. More details will be given in later sections when we systematically consider $1/m_Q$ corrections.
\section{Covariant model}
Although the development of HQET from QCD has simplified the analysis of heavy hadron physics,
many properties of hadrons, for example, their decay constants and axial coupling constants,
are still not calculable directly from QCD. To study these quantities, one unavoidably has to
use phenomenological models to describe the structures of hadrons. These include the constituent
quark model (CQM) \cite{ISGW,ISGW2}, the MIT bag model \cite{SZ,HwangMIT}, QCD sum rules \cite{BBK},
and the light-front quark model (LFQM) \cite{Jaus,Jaus1,Jaus2}. In spite of the fact that the CQM
and the MIT bag models have been widely used, results calculated from these two models are trustworthy
only for processes involving small momentum transfers. The LFQM is a relativistic quark model with
simple boost kinematics which allows us to describe physical processes with large momentum transfers.
However, this model is not a fully Lorentz covariant \cite{CCHZ}, and this defect limits its usefulness
to  matrix elements with space-like momentum transfers ($q^2 \leq 0$) only. Moreover, the LFQM is not
capable of handling the so-called Z-diagrams \cite{CCH}. In reference \cite{CCHZ}, a covariant light-front
model of heavy meson has been suggested. However the approach taken there is not systematic, and light-quark
currents are not considered. To overcome the drawbacks mentioned above, a covariant field theoretical
model has been proposed for the heavy meson bound state problem \cite{CZ,CCZ,CZ1}. This model is fully
covariant and satisfies HQS; at the same time, it retains the simplicity of the quark model picture.
This theory allows us to formulate theoretical calculations in terms of the standard Feynman diagrams.
Therefore, the lack of Z-diagrams in the ordinary LFQM is no longer a problem. Combining this model
with HQET, we can systematically study various $1/m_Q$ corrections to heavy meson properties in the
framework of the perturbative field theory.
\subsection{Basic Formalism}
In the last section, we have shown that, as $m_Q \to \infty$, the full QCD Lagrangian:
 \be
  L &=& L_Q + L_q + L_g\non \\
 &=& \bar Q~(i\gamma_\mu D^\mu - m_Q)~Q + \bar q~(i\gamma_\mu D^\mu - m_q)~q - {1\over {4}} F^{\mu\nu}_a F_{a\mu\nu},
 \en
reduced to:
\be
   L_0 = \bar h_v iv\cdot D h_v + \bar q~(i\gamma_\mu D^\mu - m_q)~q - {1\over {4}} F^{\mu\nu}_a F_{a\mu\nu}. \label{L0}
\en
This Lagrangian is responsible for binding a heavy quark and a light quark in the heavy quark limit. Since an exact solution to the QCD bound state problem does not exist, we shall take a phenomenological approach
by assuming that, after summing all the two-particle irreducible diagrams for a heavy-light system, the effective coupling between a heavy quark ($\psi_Q$) and a light quark ($\psi_q$) can be written as \cite{CZ,CCZ}:
 \be
 L_I^{Qq} = g_0 \bar{h}_v i\gamma_5 [F(-iv\cdot\partial)\psi_q] \cdot
                   [F(i\cdot\partial) \bar {\psi}_q] i\gamma_5 h_v,
 \en
in the pseudoscalar channel, where $g_0$ is a coupling constant, and $F$ is a form factor
whose presence is expected for an effective interaction resulting from the non-perturbative QCD dynamics. The functional form of $F$ will be specified later. $L_I^{Qq}$ can be considered as a generalized four-fermion coupling model \cite{NJL,BH} inspired by QCD in the heavy quark limit.
If, indeed, the above assumption is reasonable, $L_I^{Qq}$ should produce a bound state of pseudoscalar heavy meson with a physical mass $m_M$. Consequently,
the sum of all iterations of diagrams should have a pole at the reduced mass:
 \be
 \bar \Lambda \equiv \lim_{m_Q \to \infty} m_M-m_Q. \label{Lambda}
 \en
The corresponding sum of amplitudes is given by:
 \be
 A_{Qq} = g_0 F(v\cdot p') F(v\cdot p) {i\over 1-g_0\Pi_F(v\cdot k)}, \label{AQq}
 \en
where $\Pi_F$ comes from the quark loop. The existence of a pole at $v\cdot k=\bar \Lambda$ implies that $g_0 = {1/\Pi_F(\bar \Lambda)}$, and so
 \be
 A_{Qq}=\frac{-i F(v\cdot p') F(v\cdot p)}{\Pi'_F(\bar \Lambda) (v\cdot k-\bar \Lambda)+
       \Pi_F^r(v\cdot k)}, \label{AQq2}
 \en
where $\Pi_F(v\cdot k)$ has been expanded around $v\cdot k=\bar \Lambda$:
 \be
 \Pi_F(v\cdot k) = \Pi_F(\bar \Lambda) + \Pi_F'(\bar \Lambda) (v\cdot k - \bar \Lambda)
               + \Pi_F^r(v\cdot k). \label{Pif}
 \en

It is convenient to represent this $Q\bar q$ bound state
by a heavy meson field operator $\Phi$.  The corresponding
free Lagrangian is:
 \be
 L_0^M = \partial_\mu\Phi^\dagger \partial^\mu\Phi
 -m_M^2\Phi^\dagger\Phi.
 \en
To be consistent with HQET, we remove the heavy quark mass $m_Q$ from $L_0^M$
by redefining $\Phi$ as:
 \be
 \Phi(x) = \frac{1}{\sqrt{m_M}} ~e^{-im_Q v\cdot x} ~\Phi_v (x),
 \label{Phiv}
 \en
where $v$ is the velocity of the heavy meson. In terms of the new field $\Phi_v$,
and in the limit $m_Q \rightarrow \infty$, $L_0^M$ becomes:
 \be
 L_0^M=2 \Phi_v^\dag (iv \cdot\stackrel{\leftrightarrow}{\partial}-\bar \Lambda)\Phi_v
 \en
where $\stackrel{\leftrightarrow}{\partial} \equiv \frac{1}{2}
(\stackrel{\rightarrow}{\partial}-\stackrel{\leftarrow}{\partial})$.
Thus $\Phi_v$ corresponds to a particle with mass $\bar \Lambda$.

To study the structure of the pseudoscalar heavy meson, we first write down
the coupling between the heavy meson ($\Phi_v$) and its
constituent heavy ($\psi_Q$) and light ($\psi_q$) quarks:
 \be
 L_I^{M} = -G_0 \Phi_v \bar {h}_v i\gamma_5 \bar {F}(-iv\cdot \partial)\psi_q
               + h.c., \label{LIM}
 \en
where $G_0$ is the coupling constant and $\bar F$ is a vertex structure function related to the heavy meson bound state wave function on which we shall impose the constraint that the
heavy meson does not decay into $Q$ and $\bar q$ physically. Note that $\bar F$ is chosen to be independent of the residual momentum of the heavy quark so that the heavy quark charge is not spread out, otherwise, the Isgur-Wise function will not be properly normalized.

Next, we must match this meson representation with the original quark-quark interaction picture.
This is done by demanding that the heavy-light quark scattering amplitude ($A_M$) calculated is equivalent to the $A_{Qq}$ of Eq. (\ref{AQq2}). We readily obtain:
 \be
 A_M = \frac{i G_0^2 \bar {F}(v\cdot p) \bar{F}(v\cdot p')}
      {2(v\cdot k -\bar {\Lambda}_0)-G_0^2 \Pi_{\bar F}(v\cdot k)},
 \label{AM1}
 \en
where $\bar {\Lambda}_0$ is the bare reduced mass, and $\Pi_{\bar F}(v\cdot k)$ is the same as $\Pi_F(v\cdot k)$, except that the structure factor $F$ is replaced by $\bar F$.
Expanding $\Pi_{\bar F}(v\cdot k)$ around the mass shell ($v\cdot k=\bar {\Lambda}$) as in
Eq. (\ref{Pif}), we obtain:
 \be
 A_M = \frac{i G^2 \bar {F}(v\cdot p) \bar {F}(v\cdot p')}
       {2(v\cdot k -\bar \Lambda)-G^2 \Pi_{\bar F}^r(v\cdot k)},
 \label{AM2}
 \en
with $\bar \Lambda = \bar {\Lambda}_0 + \frac{1}{2} G_0^2 \Pi_{\bar F}(\bar \Lambda)$, $G = \sqrt{Z_3} G_0$, and $Z_3 = 1+\frac{1}{2} G^2 \Pi'_{\bar F}(\bar \Lambda)$,
where $Z_3$ is the wave function renormalization constant of the heavy meson field. $Z_3$ is set to zero because of the so-called compositeness condition, which originally proposed in Refs. \cite{Salam,Weinberg,Japans} and extensively applied in Refs. \cite{EI,ILL,IL,IKKSS}. From Eqs. (\ref{AM2}) and (\ref{AQq2}), it is seen that for $A_M=A_{Qq}$, we must have
$F=\bar F$, so that $\Pi_F=\Pi_{\bar F}=\Pi$ and
 \be
 G^2 = {-2/ \Pi'(\bar \Lambda)}.
 \label{G}
 \en
Hence, the matching condition for the heavy meson fixes the strength of the ($\Phi_v Qq$)-coupling vertex through Eq. (\ref{G}), which is related to the wave function normalization condition in the light-front quark model. The above discussion can be readily generalized to include
heavy vector mesons ($\Phi_v^\mu$). Analogous to Eq. (\ref{LIM}), the $\Phi_v^\mu Q q$ coupling can be written as
 \be
   L_I^{'M} = -G_0 \Phi_v^\mu \bar h_v \gamma_\mu F(-iv\cdot\partial)\psi_q
               + h.c., \label{LIMv}
 \en
where by heavy quark spin symmetry, the vertex structure function $F$ is the same for pseudoscalar and vector heavy mesons. Since the rest of the derivation is basically the same as above, we will skip the details here.

We have thus constructed a covariant representation for the structure of
heavy mesons in the heavy quark limit. The above results can also be considered as an effective
field theory for constituent quarks and heavy mesons. Within this framework, hadronic matrix elements are calculated via standard Feynman diagrams where heavy mesons always appear as external legs. The Feynman rules for this effective theory are shown in Figure 2. Figure 2(a) specifies the meson-$Q$-$q$ vertex with $\Gamma_M = i\gamma_5 (-\not\!\epsilon)$ for $M$ is the pseudoscalar (vector) meson. All the other Feynman rules are the same as in QCD and HQET. This model is simpler to work with than the ordinary light-front quark model.  Moreover, since it is fully covariant, we can use it to calculate hadronic form factors at arbitrary momentum transfers, which is not possible in the light-front quark model.
\begin{figure}
 \includegraphics*[width=5in]{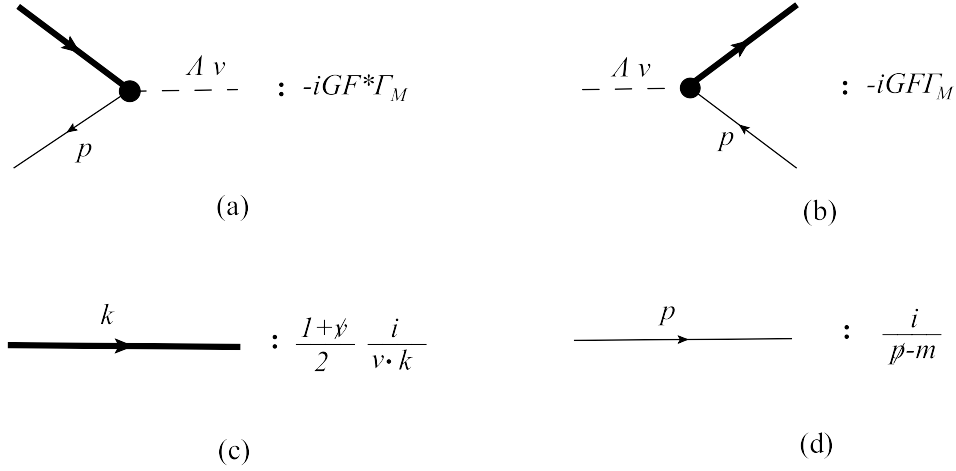}
 \caption{Feynman rules in the heavy quark limit.}
  \label{fig:2}
 \end{figure}
The self-energy of heavy meson $\Pi (v\cdot k)$ (pseudoscalar) is:
 \be
   \Pi (v\cdot k) &=& i \int \frac{d^4 p}{(2 \pi)^4} F^2(v\cdot p) \textrm{Tr}\Bigg[(i\gamma_5)\frac{-i(\not\! p -m_q)} {p^2-m^2_q}(i\gamma_5)\frac{1+\not\! v}{2}\frac{i}{(v\cdot k-v\cdot p)}\Bigg]\non \\
   &=& i \int \frac{d^4 p}{(2 \pi)^4} F^2(v\cdot p) \frac{2(v\cdot p + m_q)}{(v\cdot k-v\cdot p)(p^2-m^2_q)}.
 \en
Consequently, from Eq. (\ref{G}), we obtain:
 \be
   G^{-2} = i \int \frac{d^4 p}{(2 \pi)^4} F^2(v\cdot p) \frac{v\cdot p + m_q}{(\bar \Lambda-v\cdot p)^2(p^2-m^2_q)}. \label{Gno}
 \en
At this point, we note that $F(v\cdot p)$ is analogous to the meson wave function in the LFQM, and $G$ is the corresponding normalization constant. To explicitly evaluate $G$ and other physical quantities, we need to specify the structure function $F(v\cdot p)$, which is unfortunately not calculable from first principle.  Nevertheless, from the constraints that $F$ does not depend on the heavy quark residual momentum and it forbids on-shell dissociation of the heavy meson into $Q\bar q$, a plausible form for $F$ is
 \be
 F(v\cdot p) = \varphi(v\cdot p) (\bar \Lambda-v\cdot p), \label{FF}
 \en
where the function $\varphi(v\cdot p)$ does not have a pole at $v\cdot p = \bar \Lambda$. Accordingly, Eq. (\ref{Gno}) becomes:
 \be
   G^{-2} = i \int \frac{d^4 p}{(2 \pi)^4} \frac{|\varphi(v\cdot p)|^2(v\cdot p + m_q)}{(p^2-m^2_q)}. \label{Gno1}
 \en
Thus, if the power of $|\vec{p}|$ in the wave function $\varphi$ is less than $-\frac{2}{3}$, this model will work well.
\subsection{Heavy meson properties in the heavy quark limit}
After building a covariant framework to describe heavy meson structures, we go on to evaluate some of the basic heavy meson properties in the heavy quark limit. These include the decay constant, the axial-vector and electromagnetic coupling constants of heavy mesons.

First, consider the heavy meson decay constants defined by:
 \be
   \la 0| \bar q \gamma^\mu \gamma_5 h_v|P(v)\ra &=& i\bar f_P v^\mu, \non \\
   \la 0| \bar q \gamma^\mu \gamma_5 h_v|V(\epsilon)\ra &=& \bar f_V \epsilon^\mu. \non
 \en
The matrix element is evaluated as:
 \be
   \langle 0|\bar \psi_q \Gamma_\mu h_v|M(v)\rangle &=& -\sqrt{3} \int {d^4p\over {(2\pi)^4}}(-i)GF~ \textrm{Tr}\Bigg[{-i(\not\! p -m_q)\over {p^2-m^2_q+i\varepsilon}}\Gamma_\mu{1+\not\! v\over {2}}{i\over {\bar \Lambda -v\cdot p+i\varepsilon}}\Gamma_M\Bigg]\non \\
  &\equiv& \bar f_M~\textrm{Tr}\Bigg[{-1\over {4}}\Gamma_\mu (1+\not\! v)\Gamma_M\Bigg],
 \en
where $\sqrt {3}$ is a color factor and $\Gamma_M = i\gamma_5 (-\not\!\epsilon)$ for a pseudoscalar (vector) heavy meson; the corresponding weak current vertex is $\Gamma_\mu = \gamma_\mu\gamma_5 (\gamma_\mu)$. Thus, the decay constant in the heavy quark limit is given by:
 \be
    Tr\Bigg[{-1\over {4}}\gamma_\mu \gamma_5 (1+\not\! v)i\gamma_5\Bigg] &=& iv_\mu, \non \\
    Tr\Bigg[{-1\over {4}}\gamma_\mu (1+\not\! v)(-\not\!\epsilon)\Bigg] &=& \epsilon_\mu. \non
\en
The decay constant in the heavy quark limit is given by:
 \be
   \bar f_M = 2\sqrt{3}iG \int {d^4p\over {(2\pi)^4}} {\varphi(v\cdot p)(v\cdot p+m_q)\over {(p^2-m^2_q)}}.
 \en \label{fbm}
We find that this decay constant is the same for pseudoscalar and
vector heavy mesons, which is in accord with the prediction of HQS.
$\bar f_M$ is related to the usual definition of decay constant
$f_M$ by $f_M = \bar f_M/\sqrt{m_M}$.

Next, we study the zero order of strong coupling constants $f$ and
$g$, corresponding to $V \to P \pi$ and $V \to V \pi$, respectively.
Through PCAC, a soft pion amplitude can be related to a matrix
element of the axial vector current $A^a_\mu = \bar \psi
{\lambda^a\over {2}} \gamma_\mu \gamma_5 \psi$ as:
 \be
   \la B \pi^a (q) |A\ra = {q^\mu \over {f_\pi}} \la B |A^a_\mu | A\ra, \label{PCAC1}
 \en
From chiral Lagrangian, we obtain:
 \be
   \la P \pi^a (q) |V \ra = {-i \over {f_\pi}} {f \over {2}}~q \cdot \epsilon,  \\
   \la V \pi^a (q) |V \ra = {-i \over {f_\pi}} g \epsilon_{\mu\nu\alpha\beta} q^\mu \epsilon'^{*\nu} v^\alpha \epsilon^\beta.
 \en
On the other hand, the matrix element on the right hand side of Eq. (\ref{PCAC1}) can be evaluated in the covariant model.
The Feynman diagram to be evaluated is illustrated in Figure 2, and the relevant
matrix element is illustrated in Figure 3.
\begin{figure}
 \includegraphics*[width=2.5in]{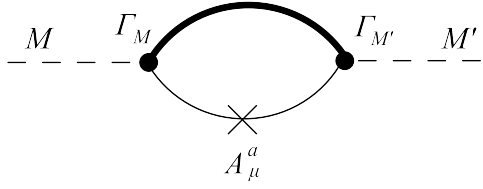}
 \caption{Feynman diagram of $f_0$ and $g_0$.}
  \label{fig:fg0}
 \end{figure}
The result is:
 \be
   \langle M'(v)|\bar \psi_q T^a\gamma_\mu \gamma_5 \psi_q|M(v)\rangle 
 &=& -\int {d^4p\over {(2\pi)^4}} [-iGF(v\cdot p)]^2 \chi^\dagger_{_{M'}} T^a \chi_{_M} \non \\
 &\times&\textrm{Tr}\Bigg[{-i(\not\! p -m_q)\over {p^2-m^2_q}}\gamma_\mu\gamma_5{-i(\not\! p -m_q)\over {p^2-m^2_q}}\Gamma_{M'}{1+\not\! v\over {2}}{i\over {\bar \Lambda -v\cdot p}}\Gamma_M\Bigg] \non \\
 &\equiv& {\cal
 G}~\textrm{Tr}\Bigg[\gamma_\mu\gamma_5\Gamma_{M'}{(1+\not\! v)\over
 {4}}\Gamma_M\Bigg]\chi^\dagger_{_{M'}} \lambda^a \chi_{_M},
 \label{fg0}
 \en
where $\chi$s are $SU(3)$ wave functions of the heavy mesons and
 \be
   {\cal G} = {-i\over {3}}G^2 \int {d^4p\over {(2\pi)^4}} |\varphi(v\cdot p)|^2 (\bar \Lambda -v\cdot p) {3m^2_q+p^2+2(v\cdot p)^2+6m_q v\cdot p\over {(p^2-m^2_q)^2}}. \label{gg}
 \en
For $V \to P\pi~(\Gamma_M = -\not\!\epsilon,~\Gamma_{M'}= i\gamma_5)$, we have:
 \be
   \textrm{Tr}\Bigg[\gamma_\mu\gamma_5\Gamma_{M'}{(1+\not\! v)\over {2}}\Gamma_M\Bigg] = 2i \epsilon_\mu.
 \en
Similarly for $V \to V\pi~(\Gamma_M = -\not\!\epsilon,~\Gamma_{M'}= -\not\!\epsilon')$,
 \be
   \textrm{Tr}\Bigg[\gamma_\mu\gamma_5\Gamma_{M'}{(1+\not\! v)\over {2}}\Gamma_M\Bigg] &=& -2i \epsilon_{\mu\nu\alpha\beta} \epsilon'^\nu v^\alpha \epsilon^\beta.
 \en
Then, comparing the above results with those obtained from the
chiral Lagrangian ${\cal L}_{VP}$ given in Eq. (\ref{LVP}), we conclude
that $f_0 = 2{\cal G}$ and $g_0 = {\cal G}$ where the subscript $0$
denotes zeroth order in $1/m_Q$. Thus, the HQS relation $f_0=2g_0$ is
satisfied in this model.

Next, we consider the coupling constants $d$ and $d''$ defined in Eq. (\ref{4d}), which govern the decay $V \to P \gamma$ and the magnetic moment of the heavy vector meson $V$, respectively. In the $m_Q \to \infty$ limit, the Feynman diagram to be calculated is similar to Figure $2.3$, except that the axial vector current $A^a_\mu$ is replaced by the light quark electromagnetic current $j_\mu = ee_q \bar \psi_q \gamma_\mu \psi_q$. The result is:
 \be
   \langle M'(v)|\bar \psi_q (iee_q\gamma_\mu) \psi_q|M(v)\rangle = -ee_q\int {d^4p \over {(2\pi)^4}} ~(-i)GF(v\cdot p) (-i)GF(v\cdot p') \non \\
 ~~~~~\textrm{Tr}\Bigg[{-i(\not\! p -m_q)\over {p^2-m^2_q+i\varepsilon}}i\gamma_\mu{-i(\not\! p' -m_q)\over {p'^2-m^2_q+i\varepsilon}}\Gamma_{M'}{1+\not\! v\over {2}}{i\over {\bar \Lambda -v\cdot p+i\varepsilon}}\Gamma_M\Bigg]\non \\
 \equiv {\cal D}e_q\textrm{Tr}\left[i\gamma_\mu\not\! q\Gamma_{M'}{1+\not\! v\over {4}}\Gamma_M \right], ~~~~~~~~~~~~~~~~~~~~~~~~~~~~~~~~~~~~~~~~~~~~~~
 \en
where $q = p'-p$. Since $q \to 0$, we can set $p=p'$ to arrive at:
 \be
 {\cal D} = 2i e G^2 \int {d^4p \over {(2\pi)^4}} |\varphi(v\cdot p)|^2(\bar \Lambda -v\cdot p)\frac{v\cdot p+m_q}{(p^2-m^2_q)^2}. \label{D11}
 \en
For $V \to P \gamma ~(\Gamma_M = -\not\!\epsilon,~\Gamma_{M'}= i\gamma_5)$,
 \be
   \textrm{Tr}\left[i\gamma_\mu\not\! q\Gamma_{M'}{1+\not\! v\over {4}}\Gamma_M \right] =i\epsilon_{\mu\nu\alpha\beta} q^\nu v^\alpha \epsilon^\beta
 \en
Similarly for $V \to V \gamma ~(\Gamma_M = -\not\!\epsilon,~\Gamma_{M'}= -\not\!\epsilon')$,
 \be
   \textrm{Tr}\left[i\gamma_\mu\not\! q\Gamma_{M'}{1+\not\! v\over {4}}\Gamma_M \right]= i( q \cdot \epsilon'~ \epsilon_\mu -  q \cdot \epsilon~ \epsilon'_\mu)
 \en
Thus, comparing the above results with those obtained from the Lagrangian ${\cal L}^{(2)}_{VP}$ given in Eq. (\ref{4d}), we obtain $d_0 = {\cal D}/2$
and $d''_0 = {\cal D}$. Again, the HQS relation $d''_0 = 2d_0$ is satisfied in this model. If we define
 \be
   d_0 = {e\over {2 m_q}}\bar d_0, \label{extract}
\en
where $\bar d_0$ is a dimensionless quantity, we can show that $\bar d_0 = -g_0$ in this model in Appendix $A$.
\subsection{$1/m_Q$ corrections}
The operators that break HQS to order $1/m_Q$ are ${\cal O}_1$ and
${\cal O}_2$ given in Eqs. (\ref{O1}) and (\ref{O2}), respectively. ${\cal
O}_1$ can be separated into a kinetic energy piece and a one-gluon
exchange piece:
 \be
   {\cal O}_1 = {\cal O}_1^{(k)} + {\cal O}_1^{(g)}
 \en
where
 \be
   {\cal O}^{(k)}_1 &\equiv& {-1\over {2m_Q}} \bar h_v [\partial_\mu \partial^\mu+ (v\cdot \partial)^2]h_v,  \\
   {\cal O}^{(g)}_1 &\equiv& {-g_s\over {2m_Q}} \bar h_v [(p+p')_\mu - v\cdot (p+p')v_\mu]A^\mu~h_v,
 \en
Also, ${\cal O}_2$ can be reexpressed as:
 \be
   {\cal O}_2 = -g_s T^a \sigma_{\mu\nu} \partial^\mu A^{a\nu}.
 \en
The Feynman rules for these HQS breaking interactions are given in Fig. 5.
 \begin{figure}
 \includegraphics*[width=6in]{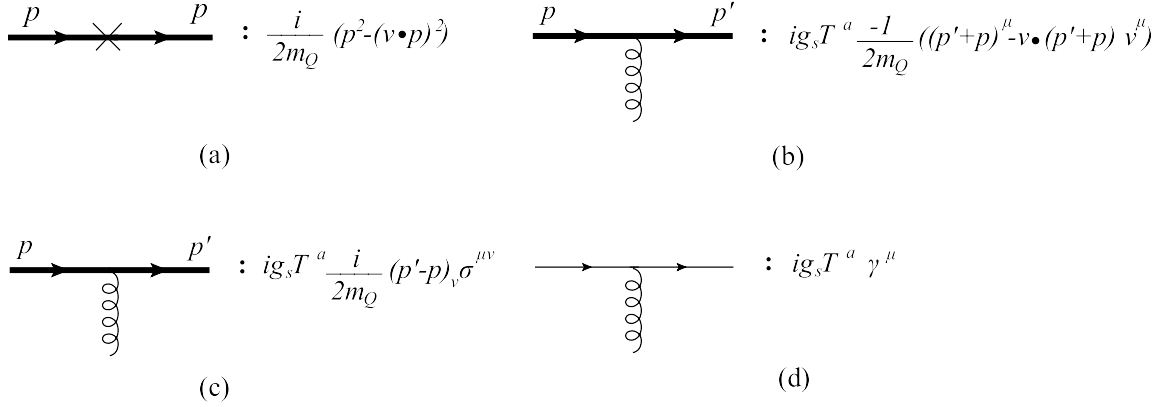}
 \caption{(a), (b), and (c) are Feynman rules for ${\cal O}^{(k)}_1$, ${\cal O}^{(g)}_1$,
 and ${\cal O}^{(g)}_2$. (d) is the light quark couples to a gluon.}
  \label{fig:3}
 \end{figure}

With the $1/m_Q$ corrections included, the heavy meson masses can be expressed as:
 \be
   m_M = m_Q + \bar \Lambda -{1\over {2 m_Q}}(\lambda_1 + d_M \lambda_2),\label{MMM}
 \en
where $\lambda_1$ comes from ${\cal O}_1$ and $\lambda_2$ comes
from ${\cal O}_2$. $\lambda_1$ receives two different contributions,
one from ${\cal O}_1^{(k)}$ and the other from ${\cal O}_1^{(g)}$,
thus:
 \be
   \lambda_1 = \lambda_1^{(k)} + \lambda_1^{(g)}.
 \en
$\lambda_1^{(k)}$ comes
from the heavy quark kinetic energy, $\lambda_1^{(g)}$ and
$\lambda_2$ are, respectively, chromoelectric and chromomagnetic
contributions. $\lambda_1$ parametrizes the common mass shift for
the pseudoscalar and vector mesons, and $\lambda_2$ accounts for the
hyperfine mass splitting. In the non-relativistic quark models, the
hyperfine mass splitting comes from a spin-spin interaction of the
form:
 \be
   H_{HF} \sim \vec S_Q \cdot \vec S_q.
 \en
where $\vec S_Q(\vec S_q)$ is the spin operator of the heavy quark (light quark). Thus:
 \be
   d_M &=& -\langle M(v) |4 \vec S_Q \cdot \vec S_q |M(v)\rangle \non \\
&=& -2[S_M(S_M+1) - S_Q(S_Q+1)- S_q(S_q+1)],
 \en
where $S_M$ is the spin quantum number of the meson $M$. Consequently, $d_M = -1$ for a
vector meson and $d_M = 3$ for a pseudoscalar meson. We will see
that this conclusion is unchanged in a relativistic framework. The
relevant Feynman diagrams are shown in Fig. 6.
\begin{figure}
 \includegraphics*[width=5in]{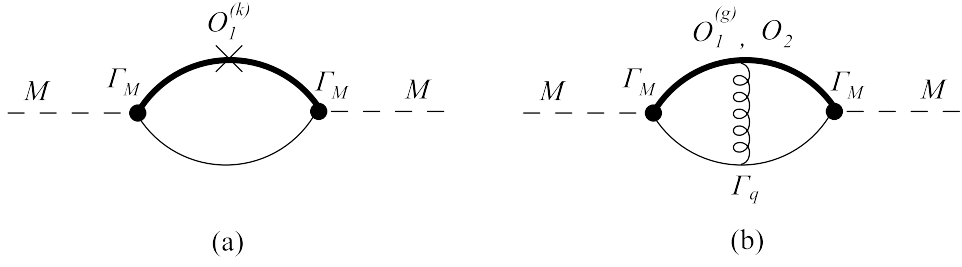}
 \caption{Feynman diagrams for $1/m_Q$ corrections to meson mass.}
  \label{fig:4}
 \end{figure}
Using Feynman rules in Figure $3$, we can readily write down the various contributions:
 \be
   {\lambda_1^{(k)} \over {2m_Q}} &=& -i\int {d^4p\over {(2\pi)^4}} [-iGF(v\cdot p)]^2   \non \\
  &\times&\textrm{Tr}\Bigg[{-i(\not\! p -m_q)\over {p^2-m^2_q}}
  \Gamma_{M}{1+\not\! v\over {2}}{i\over {\bar \Lambda -v\cdot p}}\Gamma^k_{Q1}{1+\not\! v\over {2}}{i\over {\bar \Lambda -v\cdot p}}\Gamma_M\Bigg],  \\
  {\lambda_1^{(g)} \over {2m_Q}} &=& -i\int {d^4 p~d^4 p'\over {(2\pi)^4(2\pi)^4}} (-i)GF(v\cdot p) (-i)GF(v\cdot p')~g_{\mu_\alpha}\frac{-i }{(p-p')^2}\non \\
  &\times&\textrm{Tr}\Bigg[{-i(\not\! p -m_q)\over {p^2-m^2_q}}\Gamma^\mu_q{-i(\not\! p' -m_q)\over {p'^2-m^2_q}}\Gamma_{M}{1+\not\! v\over {2}}{i\over {\bar \Lambda -v\cdot p'}}\Gamma^\alpha_{Q1}{1+\not\! v\over {2}}{i\over {\bar \Lambda -v\cdot p}}\Gamma_M\Bigg], \non \\ \\
  {d_M \lambda_2 \over {2m_Q}} &=& -i\int {d^4 p~d^4 p'\over {(2\pi)^4(2\pi)^4}} (-i)GF(v\cdot p) (-i)GF(v\cdot p')~g_{\mu_\alpha}\frac{-i }{(p-p')^2} \non \\
  &\times&\textrm{Tr}\Bigg[{-i(\not\! p -m_q)\over {p^2-m^2_q}}\Gamma^\mu_q{-i(\not\! p' -m_q)\over {p'^2-m^2_q}}\Gamma_{M}{1+\not\! v\over {2}}{i\over {\bar \Lambda -v\cdot p'}}\Gamma^\alpha_{Q2}{1+\not\! v\over {2}}{i\over {\bar \Lambda -v\cdot p}}\Gamma_M\Bigg].
 \en
where
 \be
   \Gamma^\mu_q &=& ig_s {\lambda^a \over {2}}\gamma^\mu , \non \\
   \Gamma^{(k)}_{Q1} &=& {i\over {2 m_Q}} (p^2-v\cdot p^2), \non \\
   \Gamma^\mu_{Q1} &=& ig_s {\lambda^a \over {2}}{-1\over {2 m_Q}}[(p+p')^\mu - v\cdot (p+p')v^\mu],\non \\
  \Gamma^{(g)\mu}_{Q2} &=& ig_s {\lambda^a \over {2}}{i\over {2 m_Q}}(p-p')_\nu \sigma^{\mu\nu}. \label{Gammas}
 \en
After some algebraic calculations, we obtain:
 \be
    \lambda^{(k)}_1 
    &=& i G^2\int {d^4 p\over {(2\pi)^4}} {|\varphi(v\cdot p)|^2 \over {(p^2-m^2_q)}} 2(p^2-v\cdot p^2) (v\cdot p +m_q), \label{lamonek} \\
     \lambda^{(g)}_1 
    &=&-C_f G^2 g^2_s \int {d^4 p~d^4 p'\over {(2\pi)^4(2\pi)^4}} {\varphi^\dagger(v\cdot p') \varphi(v\cdot p) \over {(p'^2-m^2_q) (p^2-m^2_q) (p-p')^2}}~{\cal T}^1_M ,\label{lamone} \\
   d_M \lambda_2 
  &=& -g_s^2 C_f G^2\int {d^4 p~d^4 p'\over {(2\pi)^4(2\pi)^4}} {\varphi^\dagger(v\cdot p') \varphi(v\cdot p) \over {(p'^2-m^2_q) (p^2-m^2_q) (p-p')^2}}~{\cal T}^2_M \label{lamtwo}
 \en
where $C_f={4 \over{3}}$ is a color factor and ${\cal T}^{1,2}_M$ 
are defined by:
 \be
   {\cal T}^1_M &\equiv& 2\{ (p\cdot p'+p'^2-v\cdot p v\cdot p'-v\cdot p'^2)(m_q+v\cdot p) \non \\
 &&+(p\cdot p'+p^2-v\cdot p v\cdot p'-v\cdot p^2)(m_q+v\cdot p')\}, \\
   {\cal T}^2_M &\equiv& {4\over {3}}d_M\{(p'^2-p\cdot p'+v\cdot p v\cdot p'-v\cdot p'^2)(m_q+v\cdot p) \non \\
 &&-(p\cdot p'-p^2-v\cdot p v\cdot p'+v\cdot p^2)(m_q+v\cdot p')\}.
 \en
where we have used the identities given in Appendix $B$. As expected, $\lambda_1^{(k)}$ and $\lambda^{(g)}_1$ are the same for both pseudoscalar and vector mesons. For convenience, we redefine:
 \be
   \lambda_1 \equiv  \lambda^{(k)}_1 + \alpha_s \bar \lambda^{(g)}_1, ~~~ \lambda_2 \equiv  \alpha_s \bar \lambda_2 \label{newdef}
 \en
where $\alpha_s = g_s^2/4\pi$. Thus, Eq. (\ref{MMM}) becomes:
 \be
   m_M = m_Q + \bar \Lambda -{1\over {2 m_Q}}(\lambda^{(k)}_1 + \alpha_s \bar \lambda_1^{(g)} + d_M  \alpha_s \bar \lambda_2), \label{MMMal}
\en
and we obtain the hyperfine mass splitting:
\be
   \Delta m_{_{HF}} = m_V -m_P = {2 \alpha_s \bar\lambda_2\over {m_Q}}. \label{masslp}
\en
A fit to the experimental value of $\Delta m_{_{HF}}$ will determine the ratio $\alpha_s/m_Q$.

Next, we move on to calculate the first order $1/m_Q$ corrections of
strong coupling constants. The relevant matrix elements are
collectively illustrated in Fig. 7. The correspondence between
Figs. $7(a)$ and $7(b)$ and various HQS breaking parameters are
tabled in Table $1$.
\begin{figure}
 \includegraphics*[width=5in]{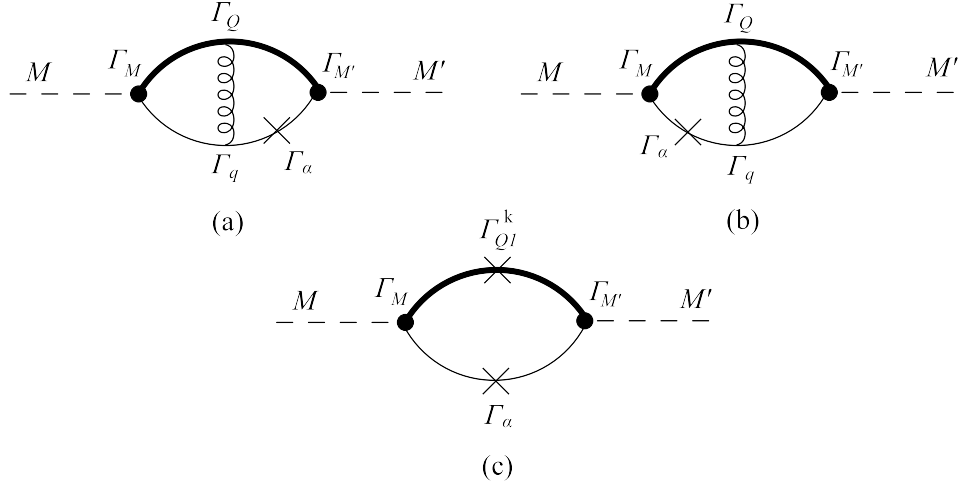}
 \caption{$1/m_Q$ corrections to strong $(f,g)$ and magnetic $(d,d'')$ coupling constants. $\Gamma_\mu$ stands for an external current. Other notations are defined in Eq. (\ref{Gammas}).}
  \label{fig:5}
\end{figure}
From experience gained in above work, we can readily write down the
matrix elements for Figures $7 (a),(b),$ and $(c)$ as:
 \be
   {\cal M}^{(a)}_\alpha &=& -i\int {d^4 p~d^4 p'\over {(2\pi)^4(2\pi)^4}} (-i)GF(v\cdot p) (-i)GF(v\cdot p')~g_{\mu_\nu}\frac{-i }{(p-p')^2} \non \\
  &\times&\textrm{Tr}\Bigg[\frac{\not\! p -m_q}{p^2-m^2_q}\Gamma^\mu_q\frac{\not\! p' -m_q} {p'^2-m^2_q}\tilde {\cal V}_\alpha\frac{\not\! p' -m_q}{p'^2-m^2_q}
  \Gamma_{M'}{1+\not\! v\over {2}}{i\over {\bar \Lambda -v\cdot p'}}\Gamma^\nu_Q{1+\not\! v\over {2}}{i\over {\bar \Lambda -v\cdot p}}\Gamma_M\Bigg], \label{dfa} \\
  {\cal M}^{(b)}_\alpha &=& -i\int {d^4 p~d^4 p'\over {(2\pi)^4(2\pi)^4}} (-i)GF(v\cdot p) (-i)GF(v\cdot p')~g_{\mu_\nu}\frac{-i }{(p-p')^2}\non \\
  &\times&\textrm{Tr}\Bigg[\frac{\not\! p -m_q}{p^2-m^2_q}\tilde {\cal V}_\alpha\frac{\not\! p -m_q}{p^2-m^2_q}\Gamma^\mu_q\frac{\not\! p' -m_q}{p'^2-m^2_q}
  \Gamma_{M'}{1+\not\! v\over {2}}{i\over {\bar \Lambda -v\cdot p'}}\Gamma^\nu_Q{1+\not\! v\over {2}}{i\over {\bar \Lambda -v\cdot p}}\Gamma_M\Bigg],  \label{dfb} \\
   {\cal M}^{(c)}_\alpha &=& \int {d^4p\over {(2\pi)^4}} [-iGF(v\cdot p)]^2 \textrm{Tr}\Bigg[\frac{\not\! p -m_q}{p^2-m^2_q}\tilde {\cal V}_\alpha \frac{\not\! p -m_q}{p^2-m^2_q}\Gamma_{M'} \non \\
  &&\times{1+\not\! v\over {2}}{i\over {\bar \Lambda -v\cdot p}} {i(p^2-v\cdot p^2) \over {2m_Q}}{1+\not\! v\over {2}}{i\over {\bar \Lambda -v\cdot p}}\Gamma_M\Bigg],  \label{dfc}
 \en
where $\Gamma_{Q} = \Gamma^{(g)}_{Q1}$ or $\Gamma_{Q2}$ and
 \be
     \tilde {\cal V}_\alpha &=& \gamma_\alpha \gamma_5 \chi^\dagger_{_{M'}} {\lambda^a \over {2}} \chi_{_M}.
 \en
\begin{table}[htb]
\begin{center}
\begin{tabular}{|c|c|c|c|l|} \hline
$\Gamma_M$ & $\Gamma_{M'}$ & $\Gamma_Q$ & $\Gamma_\mu$ & corrections \\ \hline \hline
$-\not\!\epsilon$ & $i\gamma_5$ & $\Gamma^{(g)}_{Q1}$ & $\gamma_\mu $ & $\delta f_1^{(g)} (V \to P \pi)$ \\ \hline
$-\not\!\epsilon$ & $i\gamma_5$ & $\Gamma_{Q2}$ & $\gamma_\mu $ & $\delta f_2 (V \to P \pi)$ \\ \hline
$-\not\!\epsilon$ & $-\not\!\epsilon'^*$ & $\Gamma^{(g)}_{Q1}$ & $\gamma_\mu $ & $\delta g_1^{(g)} (V \to V \pi)$ \\ \hline
$-\not\!\epsilon$ & $-\not\!\epsilon'^*$ & $\Gamma_{Q2}$ & $\gamma_\mu $ & $\delta g_2 (V \to V \pi)$ \\ \hline
\end{tabular}
\end{center}
\caption{Correspondence between Figs. $7(a)$ and $7(b)$ and $1/m_Q$ corrections to the strong coupling constants $f$ and $g$.}
\end{table}
Calculating the traces in Eqs. (\ref{dfa} $\sim$ \ref{dfc}) is very tedious but otherwise straightforward. For $V\to V \pi$ and $\Gamma_Q=\Gamma^{(g)}_{Q1}$, we obtain:
 \be
   {\cal M}^{(a)}_\alpha &=& \frac{C_f G^2}{2 m_Q} g^2_s\int {d^4 p~d^4 p'\over {(2\pi)^4(2\pi)^4}} {\varphi(v\cdot p) \varphi(v\cdot p')\over {(p^2-m^2_q)(p'^2-m^2_q)^2(p'-p)^2}} {\cal R}_1^{(a)}\label{dfga}
 \en
where
 \be
   {\cal R}_1^{(a)} &=& \Bigg\{(p'\cdot p+p^2-v\cdot p'~v\cdot p - (v\cdot p)^2)\left[(m_q + v\cdot p')^2+{1\over{3}}(p'^2-v\cdot p'^2)\right]\non \\
 &&~~+(p'^2+p'\cdot p-(v\cdot p')^2-v\cdot p' v\cdot p){2\over {3}}(m_q+ v\cdot p)(2m_q +v\cdot p')\Bigg\}.
 \en
Similarly, we can evaluate ${\cal M}^{(b)}_\alpha$, and it turns out that ${\cal M}^{(a)}_\alpha = {\cal M}^{(b)}_\alpha$. Then, a comparison with the chiral Lagrangian result shows:
 \be
   \delta g^{(g)}_1 &=& -\frac{C_f G^2}{ m_Q} \int {d^4 p~d^4 p'\over {(2\pi)^4(2\pi)^4}}
   {g^2_s\varphi^\dag(v\cdot p') \varphi(v\cdot p)
   \over {(p^2-m^2_q)(p'^2-m^2_q)^2(p'-p)^2}} {\cal R}^{(a)}_1.\label{g1g}
 \en
The above calculation can be repeated for $\Gamma_Q = \Gamma_{Q2}$.
We find that $\delta g_2$ is given by:
 \be
 \delta g_2 = \frac{C_fG^2}{2 m_Q} \int {d^4 p~d^4 p'\over {(2\pi)^4(2\pi)^4}}
 {g^2_s\varphi^\dag(v\cdot p') \varphi(v\cdot p)
 \over {(p^2-m^2_q)(p'^2-m^2_q)^2(p'-p)^2}} {\cal R}^{(a)}_2,\label{g2}
 \label{g2a}
 \en
where
 \be
   {\cal R}_2^{(a)} = \left({2\over {3}}\right)\Bigg\{\Bigg[p'^2-p'\cdot p-(v\cdot p')^2+v\cdot p' v\cdot p
   \Bigg](2m^2_q+2m_q v\cdot p +v\cdot p'~v\cdot p-p'\cdot p)\non \\
 -\Bigg[p'\cdot p-p^2-v\cdot p'~v\cdot p + (v\cdot p)^2\Bigg](m^2_q +2 m_q v\cdot p'+(v\cdot p')^2)\Bigg\}.~~~~~~~~~~~
 \en
Figure $5(c)$ corresponds to the contribution from the heavy quark kinetic energy. For $V \to V\pi$, the matrix element 
can be simplified as:
 \be
  {\cal M}^{(c)}_\alpha 
 \equiv \delta g^{(k)}_1 \textrm{Tr}\Bigg[\gamma_\alpha\gamma_5(-\not \!\epsilon'){(1+\not\! v)\over {4}}(-\not \!\epsilon)\Bigg]
 \chi^\dagger_{_{M'}} \lambda^a \chi_{_M},
 \en
where
\be
   \delta g^{(k)}_1 = i G^2 \int {d^4p\over {(2\pi)^4}} {|\varphi(v\cdot p)|^2\over {(p^2-m^2+i\epsilon)^2}}
  {(p^2-v\cdot p^2) \over {2m_Q}} \left[(m+v\cdot p)^2+{1\over {3}}(p^2-v\cdot p^2)\right].
 \label{g11k}
\en

For the process $V\to P\pi$, with coupling constant $f$, the calculation is very similar to what we have presented above.
The algebra is again rather long and tedious. Here we will only quote the results. We find that:
\be
   \delta f^{(k)}_1 = 2\delta g^{(k)}_1 ,
   \quad \delta f^{(g)}_1 = 2\delta g^{(g)}_1 ,
   \quad \delta f_2 = - 2\delta g_2 . \non
\en
These relations are identical to those found in a model independent analysis given in \cite{HYC3}. Finally we can write:
\be
   f&=& f_0 + \delta f_1^{(k)} + \delta f_1^{(g)} + \delta f_2, \non \\
   g&=& g_0 + \delta g_1^{(k)} + \delta g_1^{(g)} + \delta g_2. \non
\en

Next, we calculate $1/m_Q$ corrections to the magnetic coupling $d$ and $d''$ corresponding to $V \to P\pi$ and $V\to V\gamma$, respectively. The relevant Feynman diagrams are shown in Fig. $7$, and the correspondence between diagrams $(a)$ and $(b)$ and various HQS breaking parameters in radiative decays are listed in Table $2$.
\begin{table}[htb]
\begin{center}
\begin{tabular}{|c|c|c|c|l|} \hline
$\Gamma_M$ & $\Gamma_{M'}$ & $\Gamma_Q$ & $\Gamma_\mu$ & corrections \\ \hline \hline
$-\not\!\epsilon$ & $i\gamma_5$ & $\Gamma^{(g)}_{Q1}$ & $iee_q\gamma_\mu$ & $\delta d_1^{(g)} (V \to P \gamma)$ \\ \hline
$-\not\!\epsilon$ & $i\gamma_5$ & $\Gamma_{Q2}$ & $iee_q\gamma_\mu$ & $\delta d_2 (V \to P \gamma)$ \\ \hline
$-\not\!\epsilon$ & $-\not\!\epsilon'^*$ & $\Gamma^{(g)}_{Q1}$ & $iee_q\gamma_\mu$ & $\delta d_1^{''(g)} (V \to V \gamma)$ \\ \hline
$-\not\!\epsilon$ & $-\not\!\epsilon'^*$ & $\Gamma_{Q2}$ & $iee_q\gamma_\mu$ & $\delta d''_2 (V \to V \gamma)$ \\ \hline
\end{tabular}
\end{center}
\caption{ Correspondences between Figs. $7(a)$ and $7(b)$ and $1/m_Q$ corrections to the magnetic coupling constants $d$ and $d''$.}
\end{table}
The amplitudes for Fig. $7(a),(b)$ and $(c)$ are given by Eqs. (\ref{dfa} $\sim$ \ref{dfc}) with $\tilde {\cal V}_\alpha = iee_q \gamma_\alpha$.
For $V\to P \gamma$,
 \be
 \langle P \gamma (q,\varepsilon)|V(\epsilon)\rangle = ie_q~2d~\epsilon_{\mu\nu\alpha\beta}\varepsilon^\mu q^\nu v^\alpha \epsilon^\beta, \label{vpst1}
 \en
which come from the effective chiral Lagrangian Eq. (\ref{4d}). The calculated procedures are similar to those of the strong coupling constants. Here,
we only show the results. For $\Gamma_Q = \Gamma^{(g)}_{Q1}$,
 \be
 \delta d^{(g)}_1 &=& -\frac{2g^2_sG^2}{3m_Q}\int \frac{d^4 p~d^4 p'}{(2\pi)^8} \frac{\varphi^\dag(v\cdot p')\varphi(v\cdot p){\cal S}_1^{(a)}}{(p^2-m^2_q)(p'^2-m^2_q)^2(p'-p)^2},
 \label{dggg1}
 \en
where
 \be
  {\cal S}^{(a)}_{1} &=& -2\Bigg\{\Bigg[p'^2+p'\cdot p-(v\cdot p')^2-v\cdot p' v\cdot p\Bigg](v\cdot p +m_q) ~~\non \\
 &&~~~+\Bigg[p'\cdot p+p^2-v\cdot p'~v\cdot p - (v\cdot p)^2\Bigg](v\cdot p'+m_q) \Bigg\}.
 \en
For $\Gamma_Q = \Gamma^{(g)}_{Q2}$,
 \be
 \delta d_2 &=& \frac{2g^2_sG^2}{3m_Q}\int \frac{d^4 p~d^4 p'}{(2\pi)^8} \frac{\varphi^\dag(v\cdot p')\varphi(v\cdot p){\cal S}_2^{(a)}}{(p^2-m^2_q)(p'^2-m^2_q)^2(p'-p)^2},
 \label{dggg2}
 \en
where
 \be
  {\cal S}^{(a)}_2 &=& {4\over {3}}\Bigg\{\Bigg[p'^2-p'\cdot p-(v\cdot p')^2+v\cdot p' v\cdot p\Bigg](v\cdot p +m_q) \non \\
 &&-\Bigg[p'\cdot p-p^2-v\cdot p'~v\cdot p + (v\cdot p)^2\Bigg](v\cdot p'+m_q) \Bigg\}.
 \en
For Fig. $7 (c)$, we obtain:
\be
   \delta d_1^{(k)} =-ieG^2\int{ d^4p\over {(2 \pi)^4}} \frac{\varphi(v\cdot p)^2 (v\cdot p+m_q) (p^2-v\cdot p^2)}{2 m_Q(p^2-m_q^2)^2}. \label{dkk1}
\en
In radiative decays, there is an additional $1/m_Q$ correction which comes from the magnetic moment of the heavy quark. The matrix element of this process is:
 \be
   \langle P|\bar \psi_Q {i^2ee_Q\over {2 m_Q}}\sigma_{\mu\nu}q^\nu \psi_Q|V(\epsilon)\rangle
 =\frac{i e}{m_Q} G^2\int {d^4p \over {(2\pi)^4}} {\varphi(v\cdot p)^2 \over {(p^2-m^2_q)}}(v\cdot p +m_q) ie_Q \epsilon_{\mu\nu\alpha\beta} q^\nu v^\alpha \epsilon^\beta, \label{firstdd}
 \en
if $\Gamma_{M'} = i\gamma_5, \Gamma_{M} = -\not\!\epsilon$. From the normalization condition given in Eq. (\ref{Gno1}), we conclude that:
 \be
   \delta d_Q = {e\over {2 m_Q}}.
 \en
As to the $1/m_Q$ corrections to the radiative coupling constant $d''$ which describes the magnetic coupling $V\to V\gamma$, we will not repeat the long and tedious algebras here and just quote the final results:
 \be
   &&\delta d^{''(k)}_1 = 2 \delta d_1^{(k)}, \quad \delta d''_2 = -2 \delta d_2,\non \\
   &&\delta d^{''(g)}_1 = 2 \delta d_1^{(g)}, \quad \delta d''_Q = 2 \delta d_Q,
 \en
which again agree with the model independent analysis of Ref. \cite{HYC3}. Including the above results, we can write:
\be
   d&=& d_0 + \delta d_1^{(k)} + \delta d_1^{(g)} + \delta d_2, \non \\
   d''&=& d''_0 + \delta d_1^{''(k)} + \delta d_1^{''(g)} + \delta d''_2. \non
\en
Extracting a factor $\frac{e}{2 m_q}$ like Eq. (\ref{extract}), we can obtain the dimensionless quantities $\bar d$ and $\bar d''$.

\section{Numerical Results and Discussion}
For obtaining numerical results, we shall further assume the form of $\varphi(v\cdot p)$:
(i) $\varphi(v\cdot p)$ is an analytic function apart from isolated singularities in the
complex plane, and (ii) it vanishes as $|v\cdot p| \rightarrow \infty$. These two conditions
allow us to evaluate the $p^0-$ (or $p^--$) integrations in Eq. (\ref{Gno})
by Cauchy's Theorem. It is interesting to observe that if
 \be
 \varphi(v\cdot p) = {e^{-v\cdot p/\omega}\over \sqrt{v\cdot p+m_q}}, \label{phi}
 \en
then, the expression for decay constant $f_M$ is the same as that obtained in the covariant
light-front quark model\cite{CCHZ}, and $G$ equals the wave function normalization constant.
We note that the above expression $e^{-v\cdot p/\omega}$ is just the covariant light-front
wave function proposed in \cite{CCHZ}, while the factor 1/$\sqrt{v\cdot p+m_q}$ originates
from the Melosh transformation. However function $e^{-v\cdot p/\omega}$ is not bounded when
the four-vector $p$ is off the mass shell.  To fix this defect, one might instead want to use
$e^{-|v\cdot p|/\omega}$ or $e^{-|v\cdot p|^2/\omega^2}$ in a covariant field-theoretic formalism.
The problem is that both of these two functional forms are not analytic when continued into the
complex plane. Hence, we conclude that the widely used exponential form is not acceptable for our
purposes. For the numerical study, we shall take:
 \be
 \varphi(v\cdot p)=\frac{1}{(v\cdot p+\omega-i\varepsilon)^n} \qquad (n={\rm integer}),
 \en
which, for a sufficiently large $n$, yields very reasonable results both in the heavy quark limit
and for $1/m_Q$ corrections.

To fix the parameters ($m_q, m_Q,\omega, \alpha_s, n$) of the covariant model, we choose the data:
$f_M\simeq f_B=194 \pm 9$ MeV (an average of the results \cite{lattice1,lattice2}
in lattice QCD), $\Gamma_\textrm{tot}(D^{*+})=83.3\mp 1.8$ keV \cite{data1,data2}, and \cite{PDG12}
 \be
 r\equiv\frac{\Gamma(D^{*0}\to D^0 \gamma)}{\Gamma(D^{*0}\to D^0\pi)}=0.616\pm 0.073,\label{r1}
 \en
and follow the strategy described below. First, we choose a quark mass $m_q$ and use the above central
value of $f_B$ to determine the wave function parameter $\omega$. Subsequently we can calculate
$\lambda_1^{(k)}$, $\bar \lambda_1^{(g)}$, and $\bar \lambda_2$ from Eqs. (\ref{lamonek}) $-$ (\ref{lamtwo}).
From $\bar \lambda_2$ and the hyperfine mass splittings Eq. (\ref{HQSmass}) \cite{PDG12}, we can determine the ratio $\alpha_s/m_Q$. Now, given a value for $m_Q$, we can determine $\alpha_s$ and
use Eq. (\ref{MMM}) to obtain a corresponding $\bar \Lambda$. On one hand, knowing $m_q$, $\omega$, and $\bar \Lambda$, we can calculate $g_0$ and $d_0$ from Eqs. (\ref{gg}) and (\ref{D11}), respectively. On the other hand, with $m_Q$ and $\alpha_s$, the $1/m_Q$ corrections are included for the strong ($f$) and radiative ($d$) coupling constants from Eqs. (\ref{g1g}), (\ref{g2}), (\ref{g11k}), and (\ref{dggg1}), (\ref{dggg2}), (\ref{dkk1}), respectively. With $f$ and $d$, we can estimate $\Gamma_{\textrm{tot}}(D^{*+})$ by:
 \be
 \Gamma(V_{1/2}\to P_{1/2}+\pi^0)&=&\frac{1}{48\pi}\left(\frac{f}{\sqrt{2}f_\pi}\right)^2\left(\frac{M_P}{M_V}\right)k^3_{\pi^0}, \label{110}\\
 \Gamma(V_{1/2}\to P_{-1/2}+\pi^+)&=&\frac{1}{24\pi}\left(\frac{f}{\sqrt{2}f_\pi}\right)^2\left(\frac{M_P}{M_V}\right)k^3_{\pi^+}, \label{1-1+}\\
 \Gamma(V\to P+\gamma)&=&\frac{1}{12\pi}\mu_q^2\left(\frac{M_P}{M_V}\right)k^3_{\gamma}, \label{vpg}
 \en
where $\mu_q=2(e_q d+e_Q \frac{e}{2 m_Q})$. The value for $m_Q$ is fine-tuned to obtain a best fit to the experimental data. Finally, we repeat the above processes by fine-tuning the value for $m_q$ until the result is consistent with the data in Eq. (\ref{r1}). The results, which fit the above central values for the $D$-meson, are given in Table $3$, and those for $B$-meson are in Table $4$.
\begin{table}[htb]
\begin{center}
\begin{tabular}{|c|c|c|c|c|c|c|c|c|c|} \hline
$n$ & $m_q(\textrm{GeV})$ & $\omega(\textrm{GeV})$ & $m_c(\textrm{GeV})$ & $\alpha_s$ & $\lambda_1(\textrm{GeV}^2)$
& $\lambda_2(\textrm{GeV}^2)$ & $ \bar \Lambda(\textrm{GeV})$ & $f/2$ & $d(\textrm{GeV}^{-1})$ \\ \hline \hline
$8$ & $0.245$ & $1.47$ & $1.73$ & $0.400$ & $-0.122$ & $0.122$ & $0.210$ & $ -0.566$ & $0.361$  \\ \hline
$10$ & $0.245$ & $2.08$ & $1.72$ & $0.392$ & $-0.144$ & $0.123$ & $0.205$ & $-0.566$ & $0.361$ \\ \hline
$12$ & $0.246$ & $2.69$ & $1.72$ & $0.387$ & $-0.157$ & $0.122$ & $0.202$ & $ -0.566$ & $0.361$  \\ \hline
\end{tabular}
\end{center}
\caption{$D$-meson parameters for three different $\varphi_n$.}
\end{table}
\begin{table}[htb]
\begin{center}
\begin{tabular}{|c|c|c|c|c|c|c|c|c|c|} \hline
$n$ & $m_q(\textrm{GeV})$ & $\omega(\textrm{GeV})$ & $m_b(\textrm{GeV})$ & $\alpha_s$ & $\lambda_1(\textrm{GeV}^2)$
& $\lambda_2(\textrm{GeV}^2)$ & $ \bar \Lambda(\textrm{GeV})$ & $f/2$ & $d(\textrm{GeV}^{-1})$ \\ \hline \hline
$8$ & $0.245$ & $1.47$ & $5.09$ & $0.381$ & $-0.141$ & $0.116$ & $0.210$ & $ -0.540$ & $0.338$  \\ \hline
$10$ & $0.245$ & $2.08$ & $5.09$ & $0.373$ & $-0.162$ & $0.117$ & $0.205$ & $-0.547$ & $0.342$ \\ \hline
$12$ & $0.246$ & $2.69$ & $5.09$ & $0.368$ & $-0.175$ & $0.117$ & $0.202$ & $ -0.551$ & $0.343$  \\ \hline
\end{tabular}
\end{center}
\caption{$B$-meson parameters for three different $\varphi_n$.}
\end{table}

First of all, we see that the choices of different $\varphi_n (n=8,10,12)$ make very little difference. The value of $m_q\simeq0.245$ GeV
agrees with typical light-quark masses $m_q =0.200 \sim 0.250$~GeV used in a relativistic formulism \cite{TT}, $m_c=1.72$~GeV
and $m_b=5.09$~GeV are somewhat larger than their respective pole mass $m_c^{pole} =1.59$~GeV and $m_b^{pole}=4.89$~GeV. They are, however, consistent with $m_c=1.72 \sim 1.78$~GeV and $m_b=5.10 \sim 5.20$~GeV obtained in a Bethe-Salpeter formalism \cite{TT} and $m_b-m_c=3.37$ GeV is in agreement to that of Ref. \cite{GKLW}. Our $\lambda_1$ and $\lambda_2$ in $B$-meson are also consistent with $\lambda_1=-0.19\pm 0.10$ GeV$^2$ which is extract from CLEO data and $\lambda_2
\simeq0.12$ GeV$^2$ \cite{GKLW}, respectively. As to the reduced mass, our $\bar \Lambda$ is smaller than those obtained in  other works \cite{GKLW,GS,FL}, which is a consequence of the larger $m_Q$ used in this work.

Next, we list the $1/m_Q$ corrections to $(f,g)$ and $(\bar d,\bar d'')$ for $n=8$ in Table $5$. The parameters are $m_q=0.245^{-0.016}_{+0.019}$ GeV, $\omega=1.47^{+0.08}_{-0.10}$ GeV, $m_c=1.73^{+0.01}_{-0.02}$ GeV, and $m_b=5.09^{+0.02}_{-0.02}$ GeV. These errors come from the ones of $f_B$, $\Gamma_{\textrm{tot}} (D^{*+})$, and $r$ above.
\begin{table}[htb]
\begin{center}
\begin{tabular}{|c|c|c|c|c|c|c|c|c|} \hline
system & $-f_0$ & $ -f $ & $-g_0$ & $-g$ & $\bar d_0$ & $\bar d$ & $\bar d''_0$ & $\bar d''$  \\ \hline \hline
$D^*D$ & $1.07^{+0.01}_{-0.02}$ & $1.13^{-0.01}_{+0.02}$ & $0.534^{+0.006}_{-0.007}$ & $0.407^{+0.003}_{-0.005}$ &
$0.534^{+0.006}_{-0.007}$  & $0.583^{-0.006}_{+0.006}$ & $1.07^{+0.01}_{-0.02}$ & $0.894^{+0.014}_{-0.019}$\\ \hline
$B^*B$ & $1.07^{+0.01}_{-0.02}$ & $1.08^{+0.00}_{-0.00}$ & $0.534^{+0.006}_{-0.007}$ & $0.489^{+0.004}_{-0.006}$ &
$0.534^{+0.006}_{-0.007}$ & $0.546^{+0.002}_{-0.002}$ & $1.07^{+0.01}_{-0.02}$ & $1.01^{+0.01}_{-0.02}$ \\  \hline
\end{tabular}
\end{center}
\caption{$1/m_Q$ corrections to $(f,g)$ and $(\bar d,\bar d'')$ for $n=8$. }
\end{table}
We can see that, since $m_b/m_c \simeq 3$, consequently the HQS violating effects for the $D^*D$ system are larger than those for the $B^*B$ system by approximately a factor of $3$. For the strong decay, we have the values $\delta g_1\simeq 0.0197$ and $\delta g_2\simeq 0.0256$ in the $B^*B$ system. These lead to the ratios:
 \be
 \left(\frac{\delta g_1}{g_0}\right)=3.69\%,\qquad \left(\frac{\delta g_2}{g_0}\right)=4.79\%,
 \en
which are consistent with the rough estimate of $\alpha_s \Lambda_{\textrm{QCD}}/m_b \sim (2\sim3)\%$ for $\alpha_s \simeq 0.4$ and $\Lambda_{\textrm{QCD}} \simeq 300$~MeV. Additionally, in Refs. \cite{olddata,prc}, two strong couplings, $g_{V P\pi^0}$ and $g_{V P\pi^+}$, are defined as:
 \be
 \Gamma(V_{1/2}\to P_{1/2}+\pi^0)&=&\frac{g^2_{V P\pi^0}}{24\pi M^2_V}k^3_{\pi^0}, \\
 \Gamma(V_{1/2}\to P_{-1/2}+\pi^+)&=&\frac{g^2_{V P\pi^+}}{24\pi M^2_V}k^3_{\pi^+}.
 \en
They can be rewritten using the isospin relationship $g_{V P\pi^+}=-\sqrt{2}g_{V P\pi^+} \equiv g_{V P\pi}$ and
relating $g_{V\to P\pi}$ to a universal strong coupling between the heavy vector and pseudoscalar mesons to the pion, $\hat{g}_P$, with
 \be
 \hat{g}_P=\frac{\sqrt{2} f_\pi}{2\sqrt{M_V M_P}}g_{V P\pi}. \label{ghat}
 \en
Comparing it with Eqs. (\ref{110}) and (\ref{1-1+}), we can easily obtain $\hat{g}_P=|f/2|$. The values of $g_{V P\pi}$ and $\hat{g}_P$ compared with the experiment and other estimates are listed in Table 6.
\begin{table}[htb]
\begin{center}
\begin{tabular}{|c|c|c|c|c|} \hline
 coupling & $g_{D^* D\pi}$ & $\hat{g}_D$ & $g_{B^* B\pi}$ & $\hat{g}_B$ \\ \hline
experiment \cite{data1,data2} & $16.92 \pm 0.13 \pm 0.14$ & $0.570 \pm 0.004 \pm 0.005$ &  & \\
this work & $16.8\mp 0.2$ & $0.566^{-0.007}_{+0.008}$ & $43.9^{+0.1}_{-0.2}$& $0.540^{+0.001}_{-0.002}$  \\
DSE\cite{prc} & $15.8^{+2.1}_{-1.0}$ & $0.53^{+0.07}_{-0.03}$ & $30.0^{+3.2}_{-1.4}$ & $0.37^{+0.04}_{-0.02}$  \\
QCDSR\cite{QCDSR06} & $17.5\pm1.5$ & $0.59\pm 0.05$ & $44.7\pm1.0$ &$0.55\pm 0.01$ \\
QCDSR\cite{QCDSR01} & $14.0\pm 1.5$ & $0.47\pm 0.05$ & $42.5\pm 2.6$ & $0.52\pm 0.03$ \\
DCQM\cite{DQM} & $18\pm 3$ & $0.61\pm 0.10$ &$32\pm 5$& $0.40\pm 0.06$  \\
LQCD\cite{LQCD11} & $20\pm 2$ & $0.71\pm 0.07$ &  &   \\
LQCD\cite{LQCD09} &  &  & &$0.44^{+0.08}_{-0.03}$ \\
LQCD\cite{LQCD08} &  &  & &$0.52\pm 0.03$ \\
LQCD\cite{LQCD02} & $18.8^{+2.5}_{-3.0}$ & $0.67^{+0.09}_{-0.10}$ &  &  \\ \hline
\end{tabular}
\end{center}
\caption{Calculated values of the strong couplings compared with experiment and other estimates. (DSE: Dyson-Schwinger equation, DCQM: Dispersion approach of consistent quark model, LQCD: Lattice QCD.)}
\end{table}
Our strong couplings are consistent with those of the experiment and QCD sum rules \cite{QCDSR06}. As for the radiative decay, we use Eq. (\ref{vpg}) to calculate the decay widths $D^{*+(0)}\to D^{+(0)}+\gamma$ and $B^{*+(0)}\to B^{+(0)}+\gamma$.
The results, compared with the experiment and other estimates, are listed in Table 7. The results of LFQM \cite{choi} and the relativistic quark model (RQM) \cite{RQM} are very close to ours.
\begin{table}[htb]
\begin{center}
\begin{tabular}{|c|c|c|c|c|} \hline
 Reaction & $\Gamma(D^{*+} \to D^+\gamma)$ & $\Gamma(D^{*0} \to D^0\gamma)$ & $\Gamma(B^{*+} \to B^+\gamma)$& $\Gamma(B^{*0} \to B^0\gamma)$  \\ \hline
experiment\cite{PDG12} & $1.3\pm 0.4 $ &  &  &  \\
this work & $0.9^{+0.3}_{-0.2}$ & $22.7^{+2.1}_{-2.2}$ & $0.468^{+0.073}_{-0.075}$&$0.148\pm 0.020$  \\
LFQM\cite{choi}$^{a}$ & $0.90\pm 0.02$ & $20.0\pm 0.3$ & $0.40\pm 0.03$&$0.13\pm 0.01$  \\
RQM\cite{RQM}$^{b}$ & $0.904^{+0.025}_{-0.024}$ & $26\pm 1$ &$0.572^{+0.071}_{-0.065}$ &$0.182^{+0.022}_{-0.021}$ \\
RQM\cite{RQM1} & $1.04$ & $11.5$ &$0.19$& $0.070$  \\
LCSR\cite{LCSR} & $1.50$ & $14.40$ &$1.20$ &$0.28$  \\
HQET$+$VMD\cite{HV}& $0.51\pm 0.18$ & $16.0\pm 7.5$ &$0.22\pm 0.09$ &$0.075\pm 0.027$ \\ \hline
\end{tabular}
\end{center}
\caption{Radiative decay rates (in units of keV) in the experiment and some theoretical models. (LCSR: Light-cone sum rules, VMD: Vector meson dominance.) $^a$: The values for the linear model.$^b$: The values for $\kappa^q=0.45$.}
\end{table}

Finally, we list the predicted decay rates and branch ratios within this work and some theoretical models in Table 8. For comparison, the experimental data are also included.
{\small
\begin{table}[htb]
\begin{center}
\begin{tabular}{|c|c|c|c|c|c|} \hline
 Reaction & NR\cite{HYC4}$^{\dagger}$ & LFQM\cite{Jaus3} & SQT\cite{Rosner}& this work & experiment\cite{PDG12} \\ \hline
$D^{*+} \to D^0\pi^+$ & $78.8(68~\%)$ & $62.53(68.42~\%)$ & $(68.1\pm 0.1~\%)$ &$56.6^{-1.5}_{+1.4}(67.9\mp0.2~\%)$ & $(67.7 \pm 0.5 ~\%)$ \\
$D^{*+} \to D^+\pi^0$ & $35.7(31~\%)$ & $28.30(30.97~\%)$ & $(30.1\pm 0.1~\%)$&$25.8^{-0.7}_{+0.6}(31.0\mp0.1~\%)$ & $(30.7 \pm 0.5 ~\%)$ \\
$D^{*+} \to D^+\gamma$ & $1.9(1.7~\%)$ & $0.56(0.61~\%)$ & $(1.8\pm 0.2~\%)$&$0.9^{+0.3}_{-0.2}(1.1\pm0.3~\%)$ & $(1.6 \pm 0.4 ~\%)$ \\
$D^{*+} \to$ total & $116.4$ & $91.95$ &$80.5\pm 0.1$ &$\underline{83.3\pm 1.8}$ & $83.3\pm 1.8$\cite{data1,data2}\\ \hline
$D^{*0} \to D^0\pi^0$ & $54.1(61~\%)$ & $43.40(66.67~\%)$ &$(62.0\pm 1.7~\%)$& $36.9\mp0.9(\underline{61.9 \pm 2.9 ~\%})$ & $(61.9 \pm 2.9 ~\%)$ \\
$D^{*0} \to D^0\gamma$ & $34.0(39~\%)$ & $21.69(33~\%)$ &$(38.0\pm 1.7~\%)$ &$22.7^{+2.1}_{-2.2}(\underline{38.1 \pm 2.9 ~\%})$ & $(38.1 \pm 2.9 ~\%)$ \\
$D^{*0} \to$ total & $88.1$ & $65.09$ &$55.9 \pm1.6$ &$59.6\pm1.2$ & $< 2100$\\ \hline
\end{tabular}
\end{center}
\caption{Predicted decay rates (in units of keV) and branch ratios (in parentheses) among the four models. For comparison, the experimental branching ratios are given in the last column. (NR: Nonrelativistic quark model.) $\dagger$: The values for $g=0.5$ and $\beta=2.6$ GeV$^{-1}$.}
\end{table}}
For the total decay width of $D^{*0}$ meson, our result is close to that of LFQM \cite{Jaus3} and the single-quark-transition (SQT) formalism \cite{Rosner}.
\section{Conclusions}
In this paper, based on HQET, we have discussed the strong and radiative coupling constants of heavy mesons in $1/m_Q$ corrections symmetry breakings. These effects are studied in a fully covariant model. The covariant model starts from HQET in the heavy quark limit $(m_Q \to \infty)$, and describes a heavy meson as a composite particle, consisting of a reduced heavy quark coupled with a brown muck of light degrees of freedom. It is formulated in an effective Lagrangian approach, so that it is fully covariant, and we can use Feynman diagrammatic techniques to evaluate various processes. Especially, we have a simple relation between the strong and radiative couplings in the heavy quark limit: $\bar d_0 =-g_0$.

The parameters of this model are chosen to fit the static and decay properties of heavy mesons. For $1/m_Q$ corrections, we obtained $\delta f_2 = -2 \delta g_2$, $\delta d''_2 = -2 \delta d_2$, $\delta f^{(g,k)}_1 = 2 \delta g^{(g,k)}_1$, and $\delta d''^{(g,k)}_1 = 2 \delta d^{(g,k)}_1$, which agree with the results of a model independent analysis \cite{HYC2}. Additionally, HQS breaking effects are $3\sim 5~\%$ in magnitude for $B$ mesons, which is in accordance with the rough estimate of $\alpha_s {\Lambda_{QCD} \over{m_b}}$. Thus, we conclude that the $1/m_Q$ expansion converges very fast for bottom-hadrons. In the charmed meson sector, since $m_b/m_c \simeq 3$, consequently, the HQS violating effects are larger by approximately a factor of $3$.

For the strong couplings, our results are consistent with those of the experiment and QCD sum rules \cite{QCDSR06}. For the radiative decays, our results are close to the experimental data and the results of the light-front model \cite{choi} and the relativistic quark model \cite{RQM}. On the whole, the predictions of the total decay width of $D^{*0}$ meson in our covariant model, the light front model \cite{Jaus3} and the single-quark-transition formalism \cite{Rosner} are all in the range $\Gamma_{\textrm{tot}}(D^{*0})=55\sim65$ keV. These provide a strong vote of confidence for the validity of the covariant model.

{\bf Acknowledgements}\\
This work was supported in part by the National Science
Council of the Republic of China under Grant No.
NSC 102-2112-M-017-001-MY3.

\appendix
\section{The derivation of $\bar d_0 = -g_0$}
We analytically integrate the zero component $(p^0)$ of Eq. (\ref{gg}) and obtain:
 \be
   g_0 = -G^2\int {d^3\vec{p} \over {(2\pi)^3}} {\partial \over {\partial p^0}} \Bigg[\frac{|\varphi|^2 (\bar \Lambda -v\cdot p)}{(p^0+E_p)^2}\left((v\cdot p+m_q)^2-\frac{|\vec{p}|^2}{3}\right)\Bigg]\Bigg\vert_{p^0=E_p} . \label{a1}
 \en
Defining ${\cal A}\equiv |\varphi|^2 (\bar \Lambda -v\cdot p)/{(p^0+E_p)^2}$ and choosing $v^\mu=(1,0,0,0)$, we can reduce Eq. (\ref{a1}) as:
 \be
 g_0=-\frac{G^2}{2\pi^2} \int d |\vec{p}| |\vec{p}|^2\left[\left(\frac{\partial{\cal A}}{\partial p_0}\right)\left((p^{0}+ m_q )^2-\frac{|\vec{p}|^2}{3}\right)+2{\cal A} (p^0+m_q)\right]\Bigg\vert_{p^0=E_p}. \label{a2}
 \en
Similarly, Eq. (\ref{D11}) can be reduced as:
 \be
 \bar d_0=\frac{m_q G^2}{\pi^2}\int d |\vec{p}| |\vec{p}|^2\left[\left(\frac{\partial{\cal A}}{\partial p_0}\right)(p^{0}+m_q)+{\cal A}\right]\Bigg\vert_{p^0=E_p}. \label{a3}
 \en
From Eqs. (\ref{a2}) and (\ref{a3}), we can define $\Delta\equiv g_0+\bar d_0$ and obtain:
 \be
 \Delta=-\frac{G^2}{\pi^2} \int d |\vec{p}| |\vec{p}|^2\left[\frac{\partial{\cal A}}{\partial p_0}\frac{|\vec{p}|^2}{3}+{\cal A} p^0\right]\Bigg\vert_{p^0=E_p}.\label{a4}
 \en
Using the chain rule:
 \be
 \frac{\partial}{\partial p_0}= \frac{\partial|\vec{p}|}{\partial p^0}\frac{\partial}{\partial |\vec{p}|},
 \en
we obtain:
 \be
 \Delta=-\frac{G^2}{\pi^2} \int d |\vec{p}|\left[\frac{1}{3}|\vec{p}|^3p^0\left(\frac{\partial{\cal A}}{\partial |\vec{p}|}\right)+|\vec{p}|^2p^0{\cal A}\right]\Bigg\vert_{p^0=E_p}.\label{a5}
 \en
Finally, we take the integration by parts and obtain:
 \be
 \Delta=-\frac{G^2}{\pi^2}\left(\frac{1}{3}|\vec{p}|^3p^0{\cal A}\Bigg\vert_{p^0=E_p}\right)\Bigg\vert^{|\vec{p}|=\infty}_{|\vec{p}|=0}.
 \label{a6}
 \en
The power of $|\vec{p}|$ in the wave function $\varphi$ must be smaller than $-\frac{2}{3}$, which is the necessary condition in this model (see Eq. (\ref{Gno1})). Thus, $\Delta$ will converge to zero and $\bar d_0=- g_0$.
\section{Some Useful Formulae for Calculations in the Covariant Model}

When doing some integrals in the covariant model, there are some identities that may be useful. The derivations are asfollows:
\vskip 0.5cm
\noindent {(I)}. \normalsize $$\int d^4p~f(v \cdot p)~p^\mu = \int d^4p~f(v \cdot p)~v \cdot p~~v^\mu$$
Since:
\be
   \int d^4p~f(v \cdot p)~p^\mu = a~v^\mu \label{a},
\en
\noindent
\baselineskip 0.8cm
multiplying both sides of Eq. (\ref{a}) by $v_\mu$ gives Eq.(I).                                                                                                   \vskip 0.5cm
\noindent {(II)}. $$ \int d^4p~d^4q f(v \cdot p) f(v \cdot q) p^\mu q^\nu = \int d^4p~d^4q f(v \cdot p) f(v \cdot q) {1\over {3}}\Bigg\{\left[4v\cdot p v\cdot q-p\cdot q\right]v^\mu v^\nu $$\\
~~~~~~~~~~~~~~~~~~~~~~~~~~~~~~~~~~~~~~~$$+\left[p\cdot q-v\cdot p~v\cdot q\right]g^{\mu\nu}\Bigg\}$$
Let
\be
  \int d^4p~d^4q~f(v \cdot p)f(v \cdot q) p^\mu q^\nu = b_1~v^\mu~v^\nu + b_2~g^{\mu\nu}, \label{ab}
\en
multiplying both sides of Eq. (\ref{ab}) by $v_\mu v_\nu$, we obtain:
\be
   \int d^4p~d^4q~f(v \cdot p)f(v \cdot q) ~(v \cdot p)(v\cdot q) = b_1 + b_2. \label{b1b2}
\en
On the other hand, if we contract both sides of Eq. (\ref{ab}) with $g_{\mu\nu}$, we obtain:
\be
   \int d^4p~d^4q~f(v \cdot p)f(v \cdot q)~p\cdot q = b_1 +4b_2.  \label{b12b2}
\en
Solving Eqs. (\ref{b1b2}) and (\ref{b12b2}) for $b_1$ and $b_2$, we then reproduce Eq.(II).
\vskip 0.5cm
\noindent  {(III)}. \normalsize $$ \int d^4p~d^4q f(v \cdot p) f(v \cdot q) f(v \cdot p') p'^\alpha p^\beta q^\nu = c_1 v^\alpha v^\beta v^\nu + c_2 v^\alpha g^{\beta\nu} + c_3 v^\beta g^{\alpha\nu} + c_4 v^\nu g^{\alpha\beta}$$
\vskip 0.5cm
where $p'=p+q$ and
\be
   c_1 &=& {1\over {3}} \int d^4p~d^4q f(v \cdot p) f(v \cdot q) f(v \cdot p')~~\{v \cdot p' v \cdot p v \cdot q \non \\
&&~~~~~~~~~~~~~~~~~~~~~~~~~~~~~~~~-(v \cdot p'~p\cdot q+v \cdot p~p'\cdot q+v \cdot q~p'\cdot p)\}, \non  \\
   c_2 &=& {1\over {3}} \int d^4p~d^4q f(v \cdot p) f(v \cdot q) f(v \cdot p')~~v \cdot p'(p \cdot q-v \cdot p~v \cdot q), \non  \\
   c_3 &=& {1\over {3}} \int d^4p~d^4q f(v \cdot p) f(v \cdot q) f(v \cdot p')~~v \cdot p (p' \cdot q-v \cdot p'~v \cdot q), \non  \\
   c_4 &=& {1\over {3}} \int d^4p~d^4q f(v \cdot p) f(v \cdot q) f(v \cdot p')~~v \cdot q (p' \cdot p-v \cdot p'~v \cdot p). \non
\en
\baselineskip 0.8cm
Contracting both sides of Eq.(III) with $v_\alpha v_\beta v_\nu$, $v_\alpha g_{\beta\nu}$, $v_\beta g_{\alpha\nu}$ and $v_\nu g_{\alpha\beta}$, we obtain:
\be
   \int d^4p~d^4q f(v \cdot p) f(v \cdot q) f(v \cdot p') v \cdot p' v \cdot p v \cdot q &=& c_1+c_2+c_3+c_4, \label{c1} \\
   \int d^4p~d^4q f(v \cdot p) f(v \cdot q) f(v \cdot p') v \cdot p' p \cdot q &=& c_1+4c_2+c_3+c_4, \label{c2} \\
   \int d^4p~d^4q f(v \cdot p) f(v \cdot q) f(v \cdot p')~~v \cdot p p' \cdot q &=& c_1+c_2+4c_3+c_4,  \label{c3}
\en
and
\be
   \int d^4p~d^4q f(v \cdot p) f(v \cdot q) f(v \cdot p') v \cdot q p' \cdot p &=& c_1+c_2+c_3+4c_4 \label{c4}
\en
respectively. Then, we can obtain Eq.(III) by solving Eqs. (\ref{c1} $\sim$ \ref{c4}) for the $c$'s.

\end{document}